\documentclass[useAMS,usenatbib]{mnras}
\usepackage[pdftex]{graphicx} 
\usepackage{hyperref}
\usepackage{xcolor} 
\usepackage{lscape}
\usepackage{calc}
\usepackage{amsmath}
\usepackage{afterpage}
\usepackage{bm}
\usepackage{amssymb}

\hypersetup{colorlinks,linkcolor={red!50!black}, citecolor={blue!50!black},urlcolor={blue!80!black} } 

\title[Protoclusters during the EoR]
{The FLAMINGO simulation view of cluster progenitors observed in the epoch of reionization with JWST}

\author[Lim et al.]
{Seunghwan Lim$^{1,2}$\thanks{E-mail: sl2207@cam.ac.uk},
Sandro Tacchella$^{1,2}$, 
Joop Schaye$^{3}$, 
Matthieu Schaller$^{4,3}$, 
Jakob M. Helton$^{5}$, 
\newauthor
Roi Kugel$^{3}$,
Roberto Maiolino$^{1,2}$
\\
\vspace*{6pt} \\
$^{1}$Kavli Institute for Cosmology, University of Cambridge, Madingley
Road, Cambridge, CB3 0HA, UK \\
$^{2}$Cavendish Laboratory, University of Cambridge, 19 JJ Thomson
Avenue, Cambridge, CB3 0HE, UK \\
$^{3}$Leiden Observatory, Leiden University, PO Box 9513, 2300 RA Leiden, the Netherlands \\
$^{4}$Lorentz Institute for Theoretical Physics, Leiden University, PO box 9506, 2300 RA Leiden, the Netherlands \\
$^{5}$Steward Observatory, University of Arizona, 933 N. Cherry Ave., Tucson, AZ 85721, USA
\
}

\begin{document} 

\pagerange{\pageref{firstpage}--\pageref{lastpage}}

\date{\today}
\pubyear{2023}

\maketitle

\label{firstpage}

\begin{abstract} 
Motivated by the recent JWST discovery of galaxy overdensities during the Epoch of Reionzation, we examine the physical properties of high-$z$ protoclusters and their evolution using the FLAMINGO simulation suite. We investigate the impact of the apertures used to define protoclusters, because the heterogeneous apertures used in the literature have limited our understanding of the population. Our results are insensitive to the uncertainties of the subgrid models at a given resolution, whereas further investigation into the dependence on numerical resolution is needed. When considering galaxies more massive than $M_\ast\,{\simeq}\,10^8\,{\rm M_\odot}$, the FLAMINGO simulations predict a dominant contribution from progenitors similar to those of the Coma cluster to the cosmic star-formation rate density during the reionization epoch. Our results indicate the onset of suppression of star formation in the protocluster environments as early as $z\,{\simeq}\,5$. The galaxy number density profiles are similar to NFW at $z\,{\lesssim}\,1$ while showing a steeper slope at earlier times before the formation of the core. Different from most previous simulations, the predicted star-formation history for individual protoclusters is in good agreement with observations. We demonstrate that, depending on the aperture, the integrated physical properties including the total (dark matter and baryonic) mass can be biased by a factor of 2 to 5 at $z\,{=}\,5.5$--$7$, and by an order of magnitude at $z\,{\lesssim}\,4$. This correction suffices to remove the ${\simeq}\,3\,\sigma$ tensions with the number density of structures found in recent JWST observations. 

\end{abstract} 


\section[intro]{Introduction}
\label{sec_intro}

Galaxies and the extensive cosmic structures observable in the contemporary Universe are postulated to have originated from minute density fluctuations that occurred immediately following the inflationary epoch in the early Universe. Their hierarchical growth over cosmic epochs has been extensively documented \citep[see][]{Mo2010}. The emergence of the inaugural galaxies and primal cosmic structures, originating within regions characterized by the highest density fluctuations, is believed to have played a pivotal role in the reionization and episodic star-formation events in the Universe \citep[e.g.,][]{Furlanetto2006, Matthee2015, Ishigaki2016, Overzier2016}. Protoclusters, identified as overdensities of galaxies in the early Universe, have been demonstrated through both theoretical frameworks and observational data to contribute significantly, to over 20 per cent of the cosmic star-formation activity until approximately the cosmic noon at redshift $z\,{\simeq}\,2$ \citep[e.g.,][]{Casey2015, Umehata2015, Chiang2017, Shi2020}. Recently, \citet{Sun2024} estimated about 50 per cent of the total star formation has occurred in protocluster environments at $z\,{\simeq}\,5$. This underscores the substantial significance of this population in comprehending the genesis of galaxies and their concurrent evolution within the broader context of large-scale cosmic environments. 

Despite their significance, protocluster studies encounter numerous critical challenges that pose difficulties in achieving a thorough, unbiased analysis and a comprehensive understanding of the population (see \citealt{Lim2021} for an extensive discussion). Firstly, protoclusters, as implied by their nomenclature, are commonly regarded as the high-redshift precursors of clusters observable at $z\,{\simeq}\,0$ \citep[e.g.,][]{Chiang2013, Overzier2016, Alberts2022}. While the theoretical definition is straightforward—identifying them as progenitors of clusters at high redshifts through the construction of merger trees from a model—observations rely on the spatial distribution of galaxies in `snapshots' of the Universe at specific temporal junctures for identification. Consequently, in observational contexts, protoclusters are typically `discovered' by discerning overdensities of galaxies of various types as tracers. These encompass `normal' galaxies such as Lyman break galaxies (LBGs; e.g., \citealt{Toshikawa2012, Toshikawa2018}) and H$\alpha$ emitters (HAEs; e.g., \citealt{Darvish2020, Helton2023b}). Another type of normal galaxies as a tracer is Ly$\alpha$ emitters (LAEs; \citealt{Ouchi2005, Venemans2007b, Lee2014, Dey2016}). Protoclusters, however, due to their extreme and rare nature, are often times searched for more efficiently by utilizing rare `signpost' galaxies such as dusty star-forming galaxies (DSFGs; e.g., \citealt{Oteo2018}) and sub-mm galaxies (SMGs; \citealt{Chapman2009, Umehata2015, Miller2018}). Other signposts proven effective in discovering protoclusters include quasi-stellar objects (QSOs; e.g., \citealt{Kashikawa2007, Mignoli2020, Kashino2023}) and high-redshift radio galaxies (HzRGs; e.g., \citealt{Venemans2004}). Recently, studies have also searched for Lyman-alpha blobs (LABs; \citealt{Overzier2016, Ramakrishnan2023}) as potential candidate regions. 

Moreover, the diverse range of tracers and selection techniques employed in protocluster discovery introduces complexities that hinder consistent comparisons across studies. This variability may introduce biases favoring specific subsets of the population, complicating efforts to faithfully replicate each identification technique for generating relevant mock sample from theoretical models. Notably, even when employing identical techniques, diverse apertures have been utilized to define protoclusters and estimate their integrated properties, as exemplified in Table~3 of \citet{Harikane2019}. These variations are typically uncorrected for during comparative analyses. Furthermore, the systematics and uncertainties inherent in the observations and assumptions used for property estimation, often challenging to reliably quantify, add to the intricacy of studying overdensities. Lastly, as the most extreme and rare entities in the Universe, the realization of large, well-defined representative sample of protoclusters remains an unmet challenge in both observational and hydrodynamical simulation domains, primarily due to limitations imposed by small survey volumes and simulation box sizes, respectively \citep[e.g.,][]{Lim2021, Ramakrishnan2023, Helton2023b}. 

Fortunately, recent developments mark a significant evolution in the landscape. On the observational front, there has been a notable surge in the identification of protocluster candidates at exceptionally high redshifts ($z\,{\gtrsim}\,5$), as evidenced by recent studies \citep{Harikane2019, Castellano2023, Kashino2023, WangF2023, Brinch2024}. This progress has been primarily made possible thanks to advanced observing facilities and surveys, notably the James Webb Space Telescope (JWST) and the Systematic Identification of LAEs for Visible Exploration and Reionization Research Using Subaru HSC (SILVERRUSH; \citealt{Ouchi2018}). The recent discovery with JWST, in particular, revealed the earliest galaxy overdensities to date \citep{Laporte2022, Arribas2023, Morishita2023, Scholtz2023, Tacchella2023}. While previous protocluster identifications were often undertaken individually, certain recent studies leveraging these observations have conducted systematic searches within well-defined fixed volumes \citep{Higuchi2019, Helton2023b, Brinch2023}, facilitating unbiased statistical analyses of their number densities for the first time. From a theoretical standpoint, increased computing power has enabled hydrodynamical simulations with significantly larger box sizes, up to a few Gpc, capable of encompassing a substantial number (exceeding a few thousand) of the most massive cosmic structures while maintaining the resolution to depict galaxies akin to our Milky Way \citep{Schaye2023}. It also complements and enables comparisons with predictions from a combination of N-body simulations and semi-analytic models, which has proven useful for providing guidelines for observations \citep[e.g.][]{Chiang2013, Chiang2017}. The confluence of these observational and theoretical advancements presents a timely and inaugural opportunity to undertake studies aimed at achieving a robust and comprehensive understanding of the protocluster population. 

In this study, we investigate the physical properties of protoclusters and their evolution over the time period spanning from redshift of 8, i.e. before or the onset of reionization epoch, to the present day, using the cosmological hydrodynamic simulations of FLAMINGO project \citep{Schaye2023, Kugel2023}. We particularly focus on exploring the impact of different physical scales and volumes when defining protoclusters on the estimation of their properties and projection of their fate to the present day. The fundamental discrepancy in the identification of protoclusters between observation and theory, as described above, renders the theoretical studies twofold: whether theory guides observations in how to select objects that would most likely evolve into clusters at later times, or in what is the nature of observationally selected candidates. Since the objectives of our study align with the latter perspective, we utilize a hybrid methodology to construct mock protocluster sample, which combines both the merger tree {\it and} distribution of galaxies in the snapshots, facilitating an observationally motivated selection. 

The structure of this paper is as follows. Section~\ref{sec_model} briefly describes the FLAMINGO simulation suites and the data products, and explains how we make the mock protocluster sample out of the data. Our main results, including the basic properties of our mock sample, their abundance and cosmic contribution, as well as the redshift evolution and radial profiles which can be used to predict their present-day mass and fate, are presented in Sect.~\ref{sec_result}. The results are compared with observations in Sect.~\ref{sec_obs}, highlighting a recent tension on the number density at $z\,{\gtrsim}\,5$ and the implication of our results on the tension. In Sect.~\ref{sec_discussion}, we discuss potential impacts of numerical resolution, and uncertainties associated with the subgrid models assumed in the simulation. We summarize our findings and conclude in Sect.~\ref{sec_summary}.

\section[model]{Models and methods}
\label{sec_model}

\subsection{The FLAMINGO project}
\label{ssec_FLAMINGO}

For our analysis, we adopt the data products from the Full-hydro Large-scale structure simulations with All-sky Mapping for the Interpretation of Next Generation Observations (FLAMINGO; \citealt{Schaye2023, Kugel2023}) project of the Virgo consortium. The FLAMINGO suites are hydrodynamical cosmological simulations, which improve upon BAHAMAS \citep{McCarthy2017, McCarthy2018}. One of its flagship runs has the same baryon resolution of $m_{\rm gas}\,{=}\,1.07\times10^9\,{\rm M_\odot}$ as in BAHAMAS but extends to a much bigger volume of $(2.8\,{\rm cGpc})^3$, more than 100 times bigger when compared with $(400\,h^{-1}\,{\rm cMpc})^3$ of BAHAMAS\footnote{Throughout this paper, we use ckpc, cMpc, or cGpc to refer to the comoving scales while pkpc, pMpc, or pGpc for the physical(proper) scales, unless specified otherwise.}. Another flagship simulation of the FLAMINGO project runs on a smaller box of $(1\,{\rm cGpc})^3$ but with about an order of magnitude {\it better} resolution of $m_{\rm gas}\,{=}\,1.34\times10^8\,{\rm M_\odot}$, which is the best resolution of the data products. In our study, we use both flagships runs, which we refer to as L2p8 and L1, respectively, according to the side-length of box in cMpc. The L1 suites are available at three different resolutions, spanning approximately two orders of magnitude in mass, which we will abbreviate as m8, m9, and m10, respectively (L1\_m8 or L1\_m9, for example, with m8 being the `best' resolution), based on the approximate resolution of their gas mass in logarithmic scales. In this paper, we present our results mainly based on the high-resolution runs of L1\_m8 and the large-box runs of L2p8\_m9 to best achieve both the dynamic mass range and accurate predictions at massive end, while also utilizing L1\_m9 for some results in Sect.~\ref{ssec_subgrid} where the uncertainties in subgrid models are explored. The assumed cosmology in their fiducial runs is the constraints from the Dark Energy Survey year three for a spatially flat universe \citep{Abbott2022}. However some variations including the \citet{Planck2020} cosmology as well as one with a lower $S_8$ of \citet{Amon2023}.  

FLAMINGO employs a new numerical code called \textsc{Swift} \citep{Schaller2023} as a gravity and hydrodynamics solver, and solve the equations of hydrodynamics using the smoothed-particle hydrodynamics (SPH). For modelling physical processes that occur on scales not resolved, the simulations use subgrid models as prescriptions, which are improved or succeeding largely on those developed and used successfully for the OWLS \citep{Schaye2010}, cosmo-OWLS \citep{LeBrun2014}, BAHAMAS, and EAGLE \citep{Schaye2015, Crain2015} simulations. Specifically, FLAMINGO models radiative cooling and heating rates following \citet{Ploeckinger2020}, and the pressure and temperature of the multi-phase interstellar medium following \citet{Schaye2008}. In the `m8' and `m9' suites, stars are assumed to form by conversion of gas particles with $n_{\rm H}\,{>}\,10^{-1}\,{\rm cm}^{-3}$ and pressure higher than the ISM pressure by up to 0.3\,dex into stellar particles in a stochastic way to follow the star-formation rate (SFR) of \citet{Schaye2008}, which is designed to reproduce the Kennicutt-Schmidt law. 

The simulations assume a \citet{Chabrier2003} IMF and stellar mass loss by winds and supernova (SN) feedback following \citet{Wiersma2009} and \citet{Schaye2015}, and references therein with a few modifications. The subgrid prescriptions implemented for stellar feedback including SN mainly follow stochastic injections of energy from \citet{DallaVecchia2008} but with some modifications. The black holes are seeded as in \citet{DiMatteo2008} and \citet{Booth2009}, repositioned as in \citet{Springel2005} and \citet{Booth2009}, merge as in \citet{Bahe2022}, and accrete at a modified Bondi-Hoyle rate as in \citet{Springel2005} while boosted by the factor from \citet{Booth2009}. Finally, while the fiducial model for AGN feedback adopts a model of thermal injection without jets, based on \citet{Booth2009}, a kinetic jet-like model is also tested as a variation for the flagship L1\_m9. 

Whereas some parameters in the subgrid models of FLAMINGO are fixed and constrained through theoretical and empirical considerations, many others, which are those regarding stellar and AGN feedback, are free and calibrated to the observed stellar mass function (SMF) and gas mass fraction in clusters. This is because, in the simulations that adopt a similar set of choices for subgrid prescriptions, the SMF and cluster gas fraction are known to be mainly sensitive to stellar and AGN feedback, respectively (e.g. \citealt{McCarthy2017}). The free subgrid parameters are constrained through machine learning to find the values matching the observations best (see \citealt{Kugel2023} for details), with observational errors and bias taken into account by jointly fitting them. 

While we refer the reader to the original papers for more details about the simulations and prescriptions, there are two aspects to note that are relevant and important for our analysis and thus we emphasize: first of all, there are two different approaches regarding numerical convergence, i.e. how to handle subgrid parameters when simulation resolutions are changed. One is to fix the parameter values as changing resolutions, which has been adopted in simulations like IllustrisTNG \citep{Pillepich2018}, while the other is to adjust parameters with resolutions, as in EAGLE and FLAMINGO. This is also often times referred to as weak vs. strong convergence and was introduced by \citet{Schaye2015}. While one can hope to achieve numerical convergence at a certain point as resolution increases (strong convergence), it may be natural to argue that until simulations reach resolution high enough to directly describe the relevant physical processes, specific parameter values are only valid for the given resolution thus need to be re-adjusted for another resolution (weak convergence). As we use FLAMINGO as a tool for our analysis, which follows the weak convergence approach, this means that the predictions of the most critical properties from its suites with different resolutions would be similar and stable for the physical scales that overlap. It is found that FLAMINGO reproduces the observed SMF down to masses where there are only ${\simeq}\,10$ star particles (see Fig.~8 of \citealt{Schaye2023}) in a galaxy. This is particularly important as FLAMINGO does not model the large simulated volume with high enough resolutions to resolve many low-mass galaxies with many particles. Furthermore, the calibration of parameters was performed using larger-volume hypercubes based on bigger simulation volumes for the lower-resolution runs, thus more massive objects being contained with respect to the higher-resolution runs. 

Another aspect related to our analysis of FLAMINGO is the variations in subgrid models as well as cosmology, which are only available for L1\_m9. There are a total of eight variations with different combinations of parameters regarding the strengths and prescriptions of stellar and AGN feedback, including the jet-like injection model of AGN. The parameters in the variations are obtained by carrying out the same calibration procedure as in the fiducial models as described above, but to the calibration data shifted by multiples of the expected observational errors. Likewise, there are five cosmology variations including the Planck cosmology, higher neutrino masses of up to $\Sigma m_\nu = 0.24\, {\rm eV}$, and a lower matter power of $S_8=0.766$. While we present our main results based on the fiducial runs, we include some predictions from the variations in models later in Sect.~\ref{ssec_subgrid}, to evaluate the impact of uncertainties in the models and in the observations on our results.

FLAMINGO saves and provides a total of 79 (78 for L1\_m9) snapshots, separated by $\Delta z\,{=}\,0.05$ for $z\,{\leq}\,3$, $\Delta z\,{=}\,0.25$ for $3\,{\leq}\,z\,{\leq}\,5$, and a constant interval in $\log\, a$ beyond the redshift. The data products of FLAMINGO also include full-sky lightcones with \textsc{HEALPix} maps, which may be useful for protocluster studies where contamination by projections are investigated, which, however, we only plan to do in future work and is not within the scope of this paper.

\subsection{Sample selection and estimation of their properties}
\label{ssec_sel}

From the snapshots of particle information generated by \textsc{Swift}, haloes and their substructures (subhaloes) are identified by applying the halo finder \textsc{VELOCIraptor} (VR; \citealt{Elahi2019}). VR determines the (sub)structures by an iterative Friends-of-Friends (FoF) algorithm in six-dimensional phase space, also defining their centres as the position of the particles with the greatest binding energy. Only haloes (subhaloes) with more than 32 (20) particles are identified by VR, and thus used for our analysis. 

The properties of haloes and subhaloes thus identified are calculated using the Spherical Overdensity and Aperture Processor (SOAP). Based on the membership of particles as determined by VR, SOAP provides the (sub)halo and galaxy properties estimated within apertures of varying definitions, either physical or via overdensities with respect to the mean or critical density of the Universe, and also depending on whether or not to only count gravitationally bound particles. 

Another data product of the simulation we use for analysis is the merger tree. From the (sub)haloes identified by VR, the merger trees are generated using the algorithm described in \citet{JiangL2014}, which connects (sub)haloes between snapshots by following the most bound particles. 

From the simulation catalogs, we select protoclusters based on the average halo mass of the most massive progenitor of clusters selected at $z\,{=}\,0$. We begin by identifying a cluster at the last snapshot of the simulation, namely at $z\,{=}\,0$, according to its mass. Then we trace its progenitors at a higher redshift using the merger tree, defining every object collapsing to the cluster by $z\,{=}\,0$ as a progenitor. This is what is normally done to identify and study protoclusters theoretically and for comparison with observations in the literature. We take additional steps, however, as follows. Of all clusters in each given $z\,{=}\,0$ mass bin, we calculate the average halo mass of their main (most massive) progenitor at a given higher redshift of interest. We then search for a FoF halo at the higher redshift whose mass is within 0.3\,dex of the average mass of the main progenitors. We identify the neighborhood of such halo as a protocluster corresponding to the $z\,{=}\,0$ clusters, \textit{only} if the halo is the most massive one within the sphere of radius $R_{90}(z)$, the average comoving 3D radius within which 90 per cent of $z\,{=}\,0$ member galaxies are found at each redshift. The last step of accounting for only the most massive ones within the typical size of protocluster region is to ensure that the halo identified in the above procedure is isolated and does not merge into neighboring bigger haloes to become more massive clusters at a later time. At $z\,{=}\,0$, each individual halo identified by the procedure is a cluster. 

Compared with samplings entirely based on the merger tree, our selection is rather an observationally motivated one of protoclusters, since the identification of protocluster and the prediction of its fate (e.g. whether it will become a cluster by $z\,{=}\,0$, and its final mass) are often times made based on mass in observational studies. As our focus in this study is to investigate the bias and uncertainties included in observational sample of protoclusters and their properties, rather than guiding observations to identify progenitors of a particular group of $z\,{=}\,0$ clusters (which we plan to do in future work), this selection is most suitable for our purpose. However, we find no significant impact on our results when the selection is based entirely on the merger tree, except for an increased scatter in the results, which is as expected because normally the scatter in mass evolution history at $z\,{\gtrsim}\,2$ for given $z\,{=}\,0$ mass is about an order of magnitude (\citealt{Lim2021}; L21 hereafter), much bigger than the 0.3\,dex we assumed (note that L21 also showed selections based on SFR, another observationally motivated way to identify protoclusters, result in a tight mass evolution history at $z\,{\gtrsim}\,3$ with a remarkably small scatter of 0.1\,dex). Also, we confirm that we do not find significant changes in our results when varying the scatter around the average mass, or altering the average fraction of membership to define the 3D sphere to ensure the isolation. 

For the properties of (proto)clusters, galaxies, and their associated dark matter (sub)haloes, we tested the quantities obtained from SOAP using the various radii and definitions (such as whether or not to count only bound particles), and found that our conclusions do not change with the choice. This is because any such change in the definition applies consistently from the selection through the analysis, thus picking up and tracking essentially the same progenitors over time. To maximize the dynamic range of mass we probe and the number of objects that pass the criteria of minimum particle numbers, we decide to use the properties derived from all particles identified by the 6D FoF algorithm of VR for the presentation. This is except in Sect.~\ref{sec_obs} where we compare our results with observations, because the protocluster properties from observations are normally reported in terms of estimates within spheres of given average densities such as $M_{200}$, which is defined as the total mass within spheres where the average matter density is 200 times the cosmic mean density. Finally, we distinguish between ``Fornax''-, ``Virgo''-, and ``Coma''-like clusters at $z\,{=}\,0$ by their FoF mass of $(1\,{-}\,3)\times 10^{14}\,{\rm M_\odot}$, $(3\,{-}\,10)\times 10^{14}\,{\rm M_\odot}$, and ${>}\,10^{15}\,{\rm M_\odot}$, respectively, which is relevant for some presentations of our results. We then refer to the progenitors of each type of the clusters as ``Fornax progenitors'', ``Virgo progenitors'', and ``Coma progenitors'', respectively, throughout the paper. Note that these mass criteria are close to that used in \citet{Chiang2013}, except that we define their progenitors at higher redshifts not entirely based on the merger tree, as explained above. 

\begin{figure*}
\includegraphics[width=0.9\linewidth]{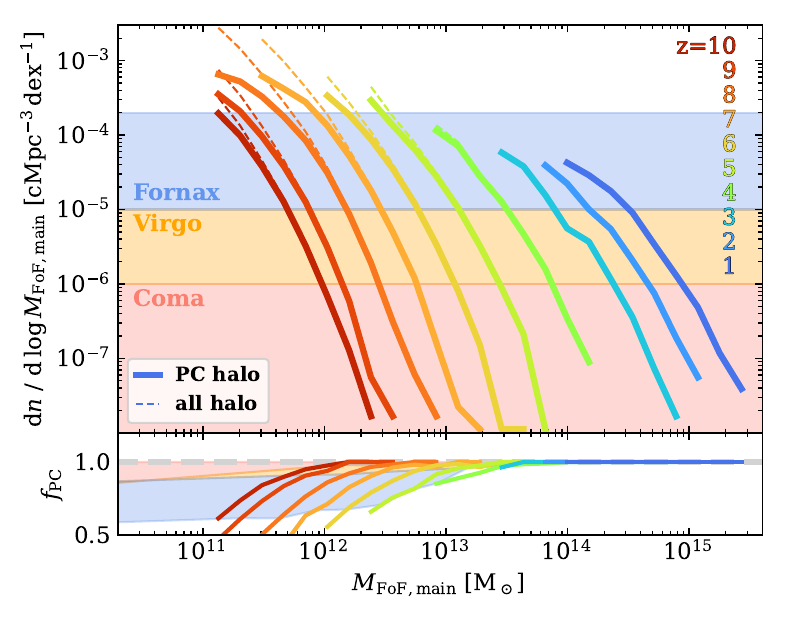}
\caption{The total mass function (the FoF mass of the most massive halo on the $x$-axis) of protoclusters selected from the FLAMINGO simulation (L1\_m8), shown by the solid lines, at various redshifts. The sample is constructed by counting all isolated haloes of mass similar to that of the average progenitor from the merger tree at each redshift (see text for more details of how we define and select the mock protoclusters). By definition, each halo thus selected at $z\,{=}\,0$ is a cluster rather than a protocluster. The shaded bands represent the approximate ranges corresponding to the three types of protoclusters, based on their $z\,{=}\,0$ mass. The dashed lines indicate the mass function of all haloes regardless of whether they are isolated (in which case we identify them as protoclusters) or have bigger haloes within the vicinity that they potentially merge into at a later time than the redshift at which the candidate haloes are investigated). The lower panel shows the ratio of isolated haloes, i.e. protoclusters, to all haloes regardless of the isolation. The lower panel indicates that up to ${\simeq}\,40$ per cent of haloes with mass matching that of the Fornax progenitors, for instance, can be misidentified as protoclusters when the identification is solely based on the halo mass.}
\label{fig_nMmain}
\end{figure*}

\section[result]{Results}
\label{sec_result}

\subsection{Basic properties of the simulated protoclusters}
\label{ssec_basic}

\begin{figure*}
\includegraphics[width=0.9\linewidth]{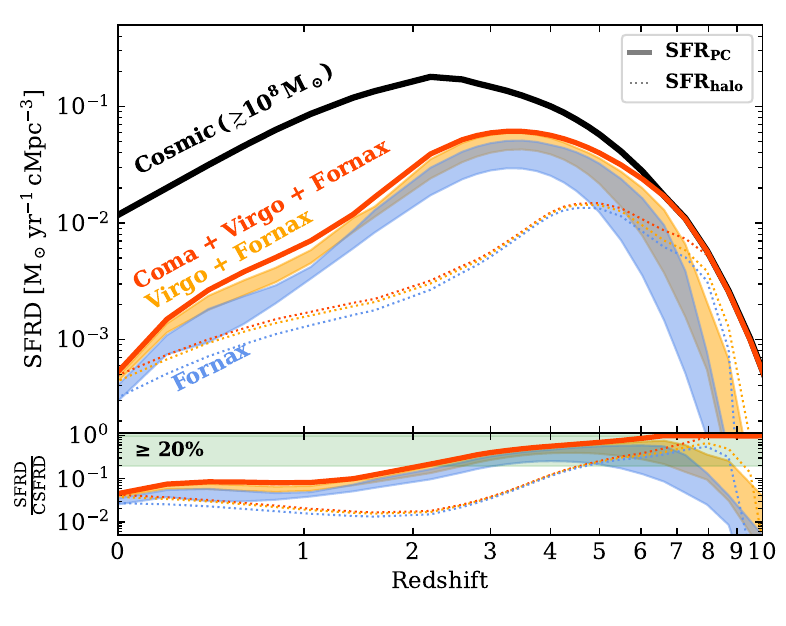}
\caption{The contribution from each type of protoclusters to the cosmic star-formation rate density (CSFRD) integrated within $R_{\rm PC}$, the typical radius of protoclusters where membership probability drops to about 50 per cent, compared with that integrated from all resolved haloes in the FLAMINGO (L1\_m8) snapshots (black solid). While the prediction for the total contribution from all three types of protoclusters is shown by the red solid line, the bands indicate the range of the predictions for the Fornax and Virgo progenitors, depending on where to put the dividing lines in the mass range to separate between the types. The SFRs are integrated down to the resolution limit of the simulation, which is the initial mean gas mass of ${\simeq}\,10^8\,{\rm M_\odot}$. The dotted lines present the same contribution but from individual haloes of similar mass to that of most massive progenitors. The lower panel shows the fraction of the contribution from each population with respect to the total cosmic SFR. }
\label{fig_CSFRD}
\end{figure*}

Here we present some basic properties of our protocluster sample from FLAMINGO. Figure~\ref{fig_nMmain} shows the comoving number densities of protoclusters from redshift of 8 to 0. $M_{\rm FoF,main}$ on the horizontal axis is the FoF mass of the most massive halo within a protocluster. The bands indicate the number densities for each of the three types of clusters, as roughly determined by the mass range of their progenitors that match their average number densities across the redshift range. We also compare those with the number density of all haloes, including those that are not the main progenitors of $z\,{=}\,0$ clusters, namely having more massive haloes at closer than $R_{90}$ that they likely merge into at a later time, instead. From the comparison, it is seen that most of the massive haloes are expected to be the main progenitors of protoclusters (see also the lower panel), being gravitational centers for cluster formation. For the haloes with mass similar to the Fornax progenitors at $z\,{\gtrsim}\,5$, however, up to about half of them merge with more massive haloes at later times (see the lower boundary of the blue shaded area in the lower panel of the figure), thus should not be considered as protoclusters. 

Next thing we investigate is the contribution from protoclusters to the cosmic star-formation rate density (CSFRD), which is another fundamental quantity for understanding the evolution of galaxies and star-formation history of the Universe. In Fig.~\ref{fig_CSFRD}, we show their SFR per comoving volume across the cosmic time separately for the Fornax, Virgo, and Coma progenitors, also comparing them with the CSFRD. For each type of protoclusters, the total SFR within $R_{\rm PC}$, the same as $R_{\rm L}$ in \citet{Chiang2017}, defined as the boundary of the protocluster where the average probability of a galaxy to be a member of the cluster at $z\,{=}\,0$ drops to 50 per cent but independently calculated for the FLAMINGO sample, is provided. The CSFRD is estimated simply by summing up the SFR contributed from all resolved ($M_\ast\,{\gtrsim}\,10^8\,{\rm M_\odot}$) haloes, and we use the instantaneous SFR from FLAMINGO as the theoretical predictions throughout the paper. Detailed comparisons with other model predictions or observational constraints may depend on the low mass limit of galaxies for which the contribution is considered, and on the timescale over which the SFR is estimated. 

Because of our selection that identifies the mock protoclusters based on the average mass and the fixed scatter of 0.3\,dex, at high redshifts where the average mass of each type of the progenitors becomes close to each other within the scatter, the same objects can be counted for more than one type. For instance, there are objects in the mass range overlap of Coma and Virgo progenitors. In order not to double-count such objects for their contribution towards each type, we provide the range of predictions by once taking those objects into Coma and once taking into Virgo type, for example, which result in the bands in the figure. The total contribution from all three types has no such ambiguity, thus indicated by the line. 

Our results show many interesting aspects. First of all, it is seen that the protoclusters, all three types combined, contribute more than 20 per cent to the CSFRD at $z\,{\gtrsim}\,2$, or for the first three billion years (see also the lower panel), confirming its importance in the evolution of galaxies and cosmic star-formation history. Also, the predicted contribution of over 50 per cent at $z\,{\gtrsim}\,5$ is in a good agreement with the estimate of \citet{Sun2024}. Secondly, the peaks are found to be shifted for protoclusters to the higher redshift of $z\,{\simeq}\,3$, relative to that for CSFRD which is around $z\,{\simeq}\,2$ as well known from observations. A similar trend was also observed by \citet{Chiang2017}, although here we find the amount of shift to be much greater. However, as can be seen later in Sect.~\ref{ssec_evol}, this is mainly due to our selection such that the number densities are not kept exactly the same across the redshifts, rather than a reflection of an intrinsic SFR evolution of individual protoclusters. 

Thirdly, most of the SFR contribution at $z\,{\lesssim}\,3$ is coming from Virgo and Fornax progenitors, rather than those of Coma. This is mainly because of the lower comoving number densities of Coma progenitors by more than an order of magnitude, as seen in Fig.~\ref{fig_nMmain}. At the higher redshifts, however, particularly at $z\,{\gtrsim}\,5$ which includes the Epoch of Reionization (EoR), the star-formation contribution from Coma progenitors dominate not only the other types of less massive protoclusters, but it basically constitutes the entire cosmic star formation. Of course at the highest redshift regime considered here, our analysis may be limited by the numerical resolution of FLAMINGO, missing a fraction of contribution from smaller ($M_\ast\,{\lesssim}\,10^8\,{\rm M_\odot}$) haloes that are not resolved and identified by the simulation. While the missing contribution from the low-mass haloes can be as large as that from the resolved haloes (see also Sect.~\ref{ssec_numres}), the fractional contribution from Coma progenitors to SFR in the early Universe is still thought to be significant, making them the most important probe of reionization and serve as places to search for and probe in order to understand the EoR. 

Lastly, we also include the prediction for the SFR contribution from all haloes (namely, regardless of whether or not isolated) with mass matching that of each type of protocluster, including those that merge into more massive haloes in the future snapshots for comparison, shown by the dotted lines. For the haloes with mass similar to that of Virgo or Fornax progenitors, such contribution appears greater than that from those identified as protoclusters at $z\,{\gtrsim}\,5$. This simply reflects the fact that there are quite a fraction of haloes that merge with bigger haloes and thus are not considered for estimating the contribution from protoclusters. At the later times of $z\,{\lesssim}\,5$ and in general, however, the contribution by protoclusters dominate that by the individual haloes of similar mass, as can be seen in the figure. This reflects the large size and comoving volume that the protoclusters occupy which includes a lot of smaller neighboring galaxies that also contribute significantly to the CSFRD.  

\begin{figure}
\includegraphics[width=0.97\linewidth]{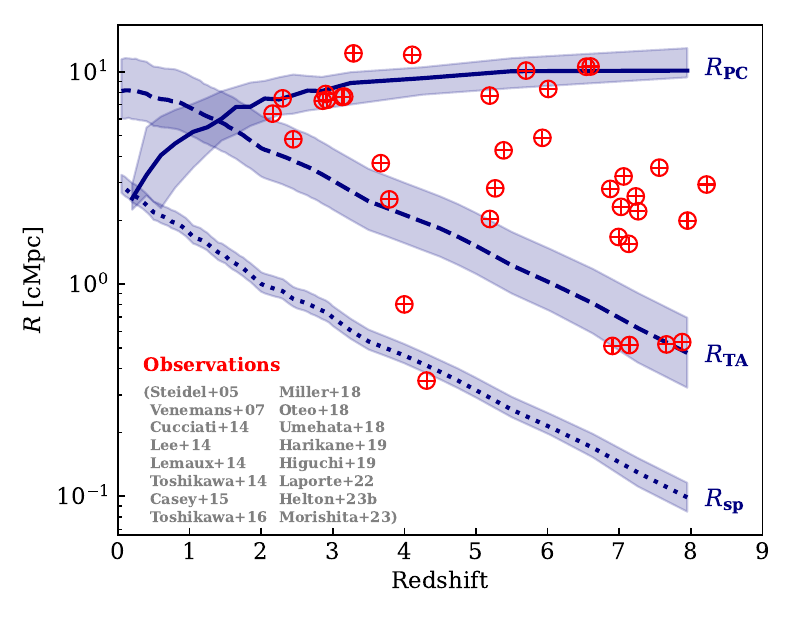}
\caption{The apertures used to define high-$z$ protoclusters selected from the literature. To highlight a large diversity uncorrelated with physically motivated apertures, they are compared with the radii for the Coma progenitors with various physical meanings such as $R_{\rm PC}$ (average radius, at each redshift, of protoclusters where membership probability drops to about 50 per cent; solid), $R_{\rm sp}$ (splashback radius defined as the apocenter of objects on their first orbit after infall; dotted), and $R_{\rm TA}$ (turn-around radius where the sphere within it is detached from the cosmic expansion and falls towards the local gravity center; dashed), estimated based on the FLAMINGO simulation (L1\_m8). See the text for details of how $R_{\rm sp}$ and $R_{\rm TA}$ were computed. The shaded areas represent the 68 per cent ranges. $R_{\rm PC}$ is therefore a theoretically motivated aperture that should be used to define and compare protoclusters, and so is $R_{\rm sp}$ for the core, while the observations are all in between, demonstrating heterogeneous aperture choices among studies. }
\label{fig_aperture}
\end{figure}

\subsection{Mass enclosed within apertures}
\label{ssec_Msph}

\begin{figure*}
\includegraphics[width=1.\linewidth]{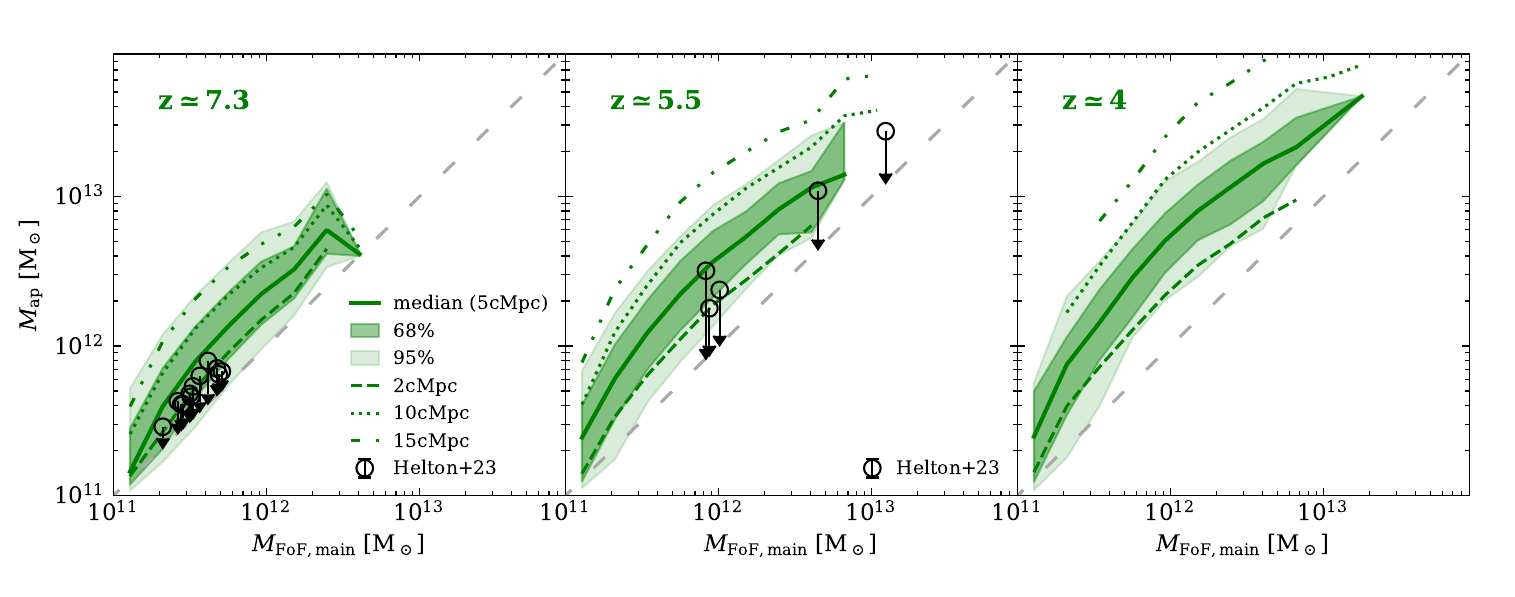}
\caption{The total FoF mass, including dark matter, of all resolved haloes enclosed within spheres of varying comoving apertures (vertical axis) versus that of the main (most massive) progenitor halo (horizontal axis) at redshifts of 7.3 (left panel), 5.5 (middle), and 4 (right). The median and 68 and 95 per cent ranges are shown for 5\,cMpc aperture (solid lines and shaded regions, respectively), while only the medians are shown for 2 (dashed), 10 (dotted), and 15\,cMpc (double dot-dashed) apertures. The recent JWST observations from \citet{Helton2023b} are included in the left and middle panels for comparison, with the circles representing the mass reported in the original paper, while the tip of the arrow representing the corresponding mass of the core estimated by correcting for the aperture. As they assumed various apertures to identify the individual protoclusters, the mass of main progenitors was estimated using the relations from FLAMINGO (L1\_m8) for each relevant aperture via interpolation. The typical bias when not corrected for apertures is about a factor of 1.5--2 at $z\,{\simeq}\,7.3$, and 3 at $z\,{\simeq}\,5.5$. }
\label{fig_Mmain_Msph}
\end{figure*}

While we have identified and so far examined protoclusters according to the mass of their most massive halo (or `core' in observations), protocluster sample and their properties from observational studies are usually identified and estimated accounting for galaxies within much bigger regions surrounding the core. Such arbitrary and heterogeneous choice of apertures are illustrated in Fig.~\ref{fig_aperture}, which we reproduced largely based on Table 3 of \citet{Harikane2019} using only those with more than 10 spectroscopically confirmed members, but with additional data from more recent observations including \citet{Laporte2022}, \citet{Helton2023b}, and \citet{Morishita2023}. The full reference for the observational data shown in the figure are \citet{Steidel2005}, \citet{Venemans2007a}, \citet{Cucciati2014}, \citet{Lee2014}, \citet{Lemaux2014}, \citet{Toshikawa2014}, \citet{Casey2015}, \citet{Toshikawa2016}, \citet{Miller2018}, \citet{Oteo2018}, \citet{Umehata2018}, \citet{Harikane2019}, \citet{Higuchi2019}, \citet{Laporte2022}, \citet{Helton2023b}, and \citet{Morishita2023}. When the aperture is only provided in angular size in the original studies, we assumed the same cosmology as FLAMINGO to convert to the comoving size. In the same figure, the observational apertures are also compared with $R_{\rm PC}$ (solid), so-called `turn-around' radius, $R_{\rm TA}$, and splashback radius, $R_{\rm sp}$, each computed for a typical Coma progenitor. The turn-around radius is the boundary where the matter inside it deviates from the expansion of the Universe and falls towards the local center of gravity, which in our case is the center of protoclusters or most massive haloes. We calculate $R_{\rm TA}$'s by adopting its ratio to $R_{\rm vir}$ of the most massive haloes from \citet{LeeJ2023}, which is typically around 4 only with a moderate evolution with time. The splashback radius, on the other hand, defined as the apocenter of objects on their first orbit after infall, is very close to $R_{\rm vir}$ for such massive objects as Coma progenitors, being about 1.2\,$R_{\rm vir}$ at $z\,{=}\,0$ and decreasing with redshift to approach unity. We particularly use the results from \citet{More2015} to deduce $R_{\rm sp}$'s for the simulated protocluster sample. As can be seen, the observational apertures spread wide ranging from $R_{\rm sp}$ to above $R_{\rm PC}$ demonstrate a great dispersion adopted among studies, potentially causing bias in the results. 

Such heterogeneous aperture choice is one of the main hurdles for analyzing protoclusters and extracting the maximum physical information required to understand their contribution to the galaxy formation and evolution. Despite its potentially important impact, however, it has been fairly neglected so far in the studies, without any correction or consideration for it in most cases when comparing samples selected from different data and inferring their physical properties. Therefore, it is important to make sure that we understand the neglected impact, which is the focus of our study. 

As a first test of such, in Fig.~\ref{fig_Mmain_Msph}, we compare the total FoF mass of all resolved haloes enclosed within various 3D apertures, with that of main halo at three selected high redshifts of 7.3, 5.5, and 4, which are near the redshifts newly discovered protoclusters from observations have piled up in recent years. We extend the aperture up to 15\,cMpc as that is about the maximum size of protocluster regions that collapse to form clusters by $z\,{=}\,0$. That also spans the range of apertures adopted in the literature. As can be seen, the impact of using different apertures on the estimation of total mass can be easily a factor of 2 to 3 bias and/or uncertainties, and up to more than an order of magnitude, for the same protocluster region with the main halo of the same mass. Therefore, if this factor is not corrected appropriately for apertures when comparing with other samples or with model predictions, the estimation of properties and their projected evolution to the present day will be significantly biased. In the figure, we also include some recent observations from \citet{Helton2023b}, who identified seventeen protocluster candidates at $5\,{\lesssim}\,z\,{\lesssim}\,9$ by pre-selecting high-$z$ galaxy candidates from the JWST Advanced Deep Extragalactic Survey (JADES; \citealt{Eisenstein2023}) and the JWST Extragalactic Medium-band Survey (JEMS; \citealt{Williams2023}), and then confirming via spectroscopy data from the First Reionization Epoch Spectroscopic COmplete Survey (FRESCO; \citealt{Oesch2023}). The candidates were traced with H$\alpha$ and [OIII]$\lambda 5008$ lines detections for the spectroscopic confirmation, which cover the redshift range of $4.9\,{<}\,z\,{<}\,6.6$ and $6.7\,{<}\,z\,{<}\,8.9$, respectively, given the wavebands of survey. They identified these candidates using a FoF alogrithm with fixed linking parameters for projected distance and line-of-sight (LOS) velocity dispersion, and then confirmed them as protocluster candidates only if the galaxy number density calculated within the aperture exceeds a certain value. The galaxy membership and mass were also estimated accordingly. However, they found that varying the linking parameters by a factor of few do not alter strongly the identification and their results. Among their seventeen samples, only eight candidates with redshifts close to those in our presentation are shown, namely, JADES-GN-OD-5.191, JADES-GN-OD-5.194, JADES-GN-OD-5.269, JADES-GN-OD-5.386, JADES-GN-OD-5.928, JADES-GN-OD-7.561, JADES-GN-OD-7.954, and JADES-GN-OD-8.220. We use the aperture and their total mass estimates within the apertures from \citet{Helton2023b}, to infer the mass of their main haloes based on the relation between the two quantities from our results. It is the mass of main halo that is normally explored in simulations and theoretical models as a proxy for predicting the fate and $z\,{=}\,0$ mass of protoclusters, while observational protoclusters are often compared to the theoretical studies using their total mass enclosed within arbitrary choice of apertures. As can be seen in the comparison with the observations from Fig.~\ref{fig_Mmain_Msph}, such estimates can be biased by a factor of 2 at $z\,{\simeq}\,7.3$, and up to 4 at $z\,{\simeq}\,5.5$, which can, in turn, also over-estimate their present-day mass and thus lead to misidentification of protoclusters, when a proper correction is not applied. The amount of correction needed for given aperture in comoving scale is even greater at lower redshifts, increasing with time, as expected from growth of structure.

\subsection{The evolution of protocluster properties}
\label{ssec_evol}

\begin{figure*}
\includegraphics[width=1.\linewidth]{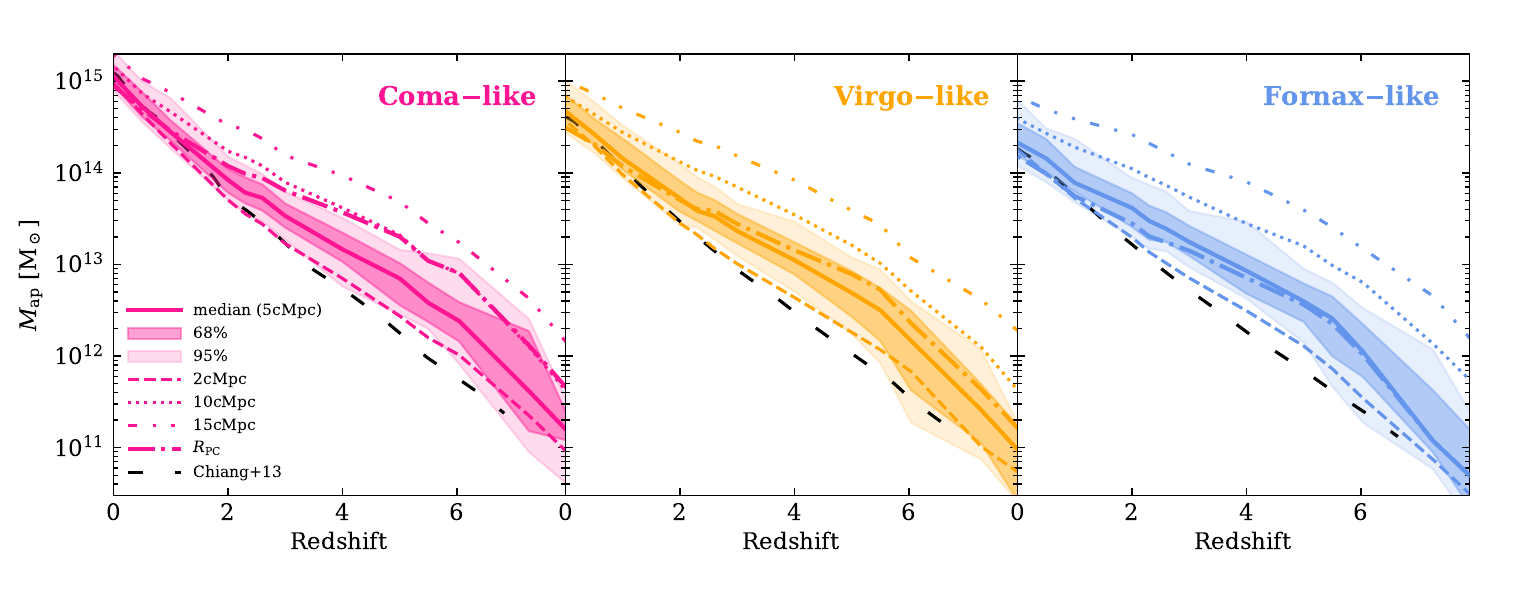}
\caption{The evolution of total mass (including dark matter, stellar and gas mass) of galaxies enclosed within varying comoving apertures for each of the three types of progenitors from the FLAMINGO (L1\_m8) simulation. Note that this is the mass of protoclusters selected independently at each snapshot, rather than tracking the mass evolution of the same objects, because of our selection as described in Sect~\ref{ssec_sel}. The median and 68 and 95 per cent ranges are shown for 5\,cMpc aperture (solid lines and shaded regions, respectively), while only the medians are shown for 2 (dashed), 10 (dotted), 15\,cMpc (double dot-dashed), and $R_{\rm PC}$ (dot-dashed) apertures. The mass of main progenitors from the simulation of \citet{Chiang2013} (black dashed) is also shown as a reference. Protoclusters increase their mass by roughly four orders of magnitude from redshift of 8 to 0. }
\label{fig_Mh_evol}
\end{figure*}

\begin{figure*}
\includegraphics[width=1.\linewidth]{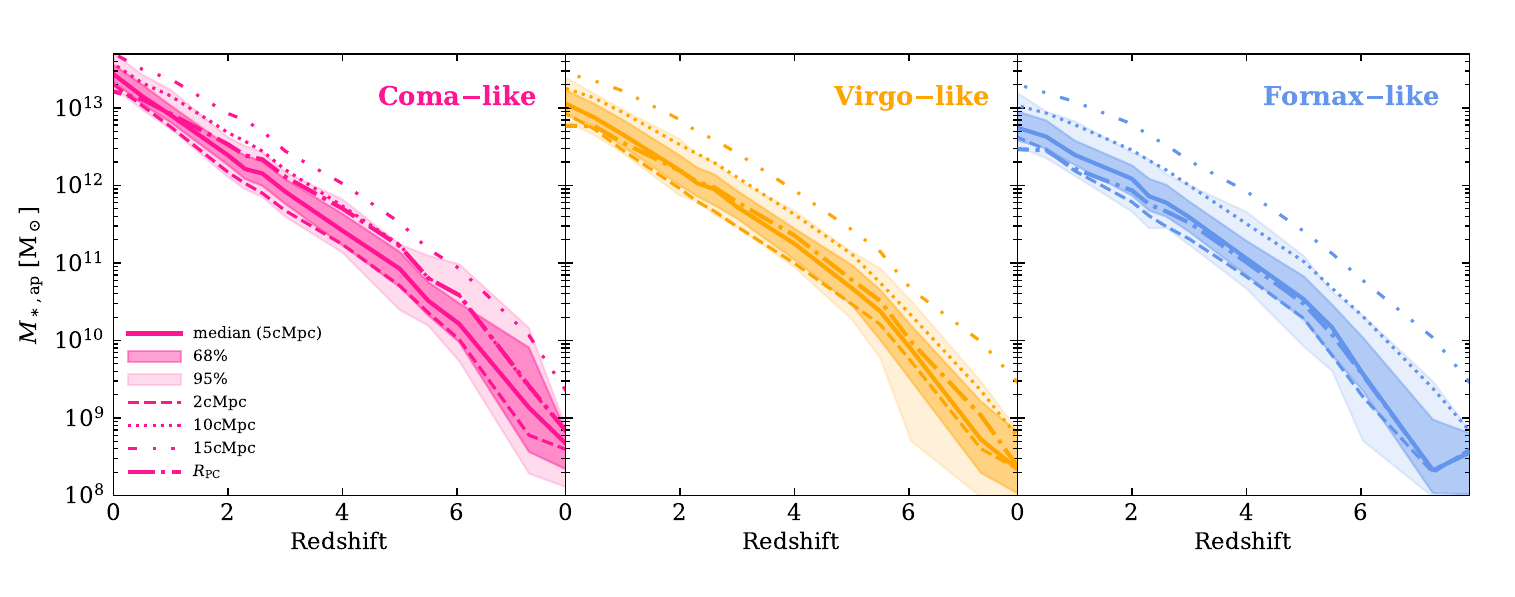}
\caption{The evolution of stellar mass of galaxies enclosed within varying comoving apertures for each of the three types of progenitors from the FLAMINGO (L1\_m8) simulation. The line styles and shadings are the same as in Fig.~\ref{fig_Mh_evol}.}
\label{fig_Ms_evol}
\end{figure*}

While we investigated how the total mass estimated for protoclusters is affected by the choice of apertures at selected redshifts, here in this section we explore the evolution of various properties for each types of protoclusters across the whole range of redshift from $z\,{=}\,8$ to the present day. Each of Coma, Virgo, and Fornax progenitors are defined and selected the same way as described in Sect.~\ref{ssec_sel}, i.e. the most massive halo within $R_{\rm PC}$ with mass matching that of progenitors of $z\,{=}\,0$ clusters. First of all, Fig.~\ref{fig_Mh_evol} shows the evolution of total mass enclosed within the same apertures as in Sect.~\ref{ssec_Msph} but with an additional aperture of $R_{\rm PC}$. For the aperture of 5\,cMpc, the 68 and 95 per cent ranges of the evolution are also provided. Our results show that Coma progenitors typically begin with seed haloes of ${\simeq}\,10^{11}\,{\rm M_\odot}$ at $z\,{\simeq}\,8$, and then evolve rapidly by accreting and accumulating mass into $R_{\rm PC}$ between redshift of 8 and 4. After redshift of 4, $R_{\rm PC}$ and the concentration of mass shrinks gradually and slowly to collapse and form the Coma-like clusters by $z\,{=}\,0$. Note that the outer envelope continues to increase in mass at lower redshifts, accreting more mass toward the already collapsed, virialized clusters at the center through $z\,{=}\,0$, by which Coma will further evolve to become bigger clusters. Similar evolutionary trends are observed for Virgo and Fornax progenitors, except that their sizes ($R_{\rm PC}$) are typically smaller and begin to shrink at earlier times of $z\,{\simeq}\,5$ and collapse much more gradually than Coma progenitors. Our results and the plots can serve as a basis for interpolation to predict the true, unbiased fate and mass of identified protoclusters candidates, given observations with the aperture and mass within it, easily applicable to observations from past studies for the corrections. 

\begin{figure*}
\includegraphics[width=1.\linewidth]{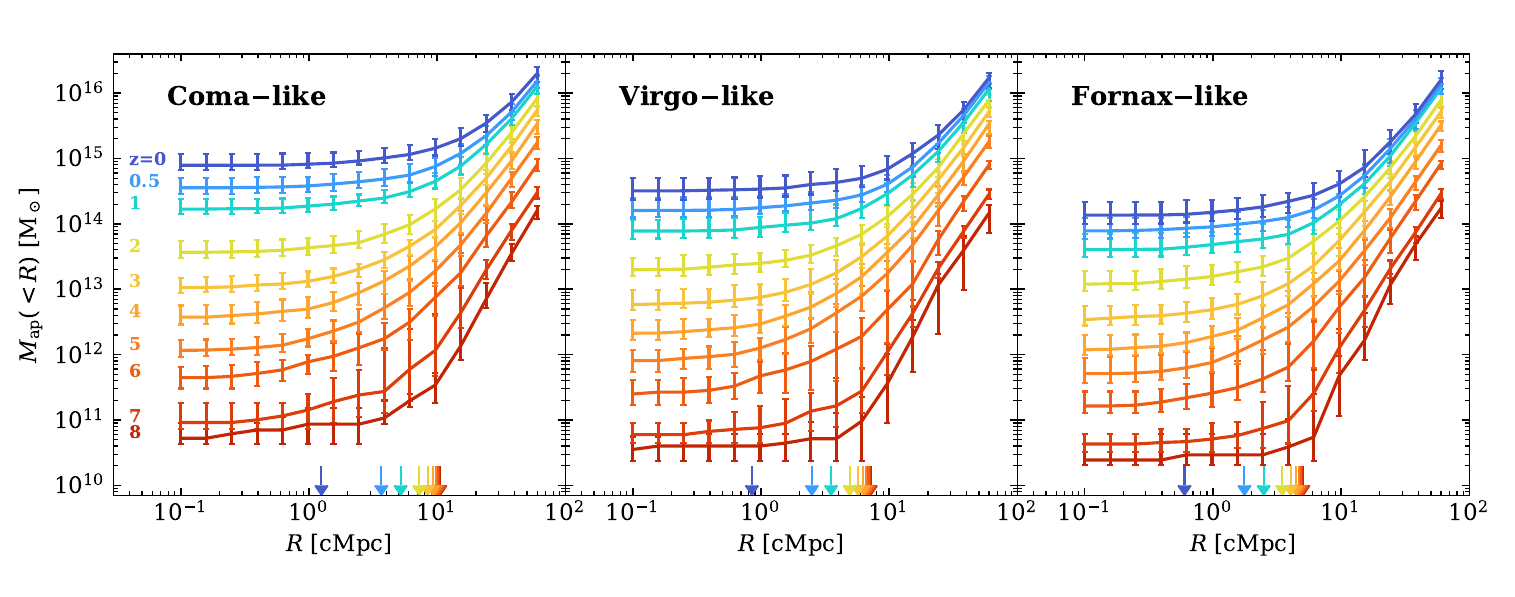}
\caption{The cumulative radial profiles of total mass as a function of comoving aperture and redshift for the three types of progenitors from the FLAMINGO (L1\_m8) simulation. The medians and the 68 per cent ranges are indicated by the solid lines and the errorbars. The values of $R_{\rm PC}$, the sizes of typical protoclusters for each type of progenitors, are indicated by the arrows at the bottom of each panel. The profiles are flattened in the innermost regions as they are dominated by the mass of the central galaxies in the protocluster `core'. }
\label{fig_Mh_prof}
\end{figure*}

\begin{figure*}
\includegraphics[width=1.\linewidth]{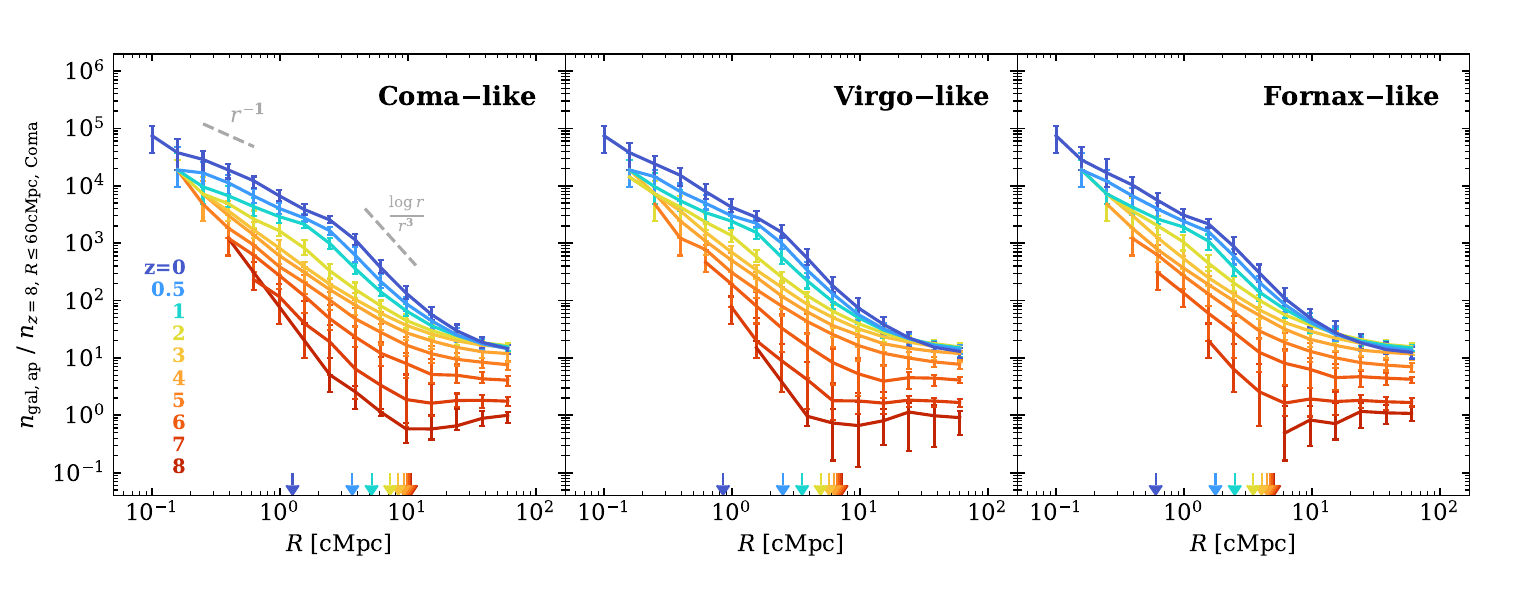}
\caption{The cumulative radial profiles of galaxy number density, normalized to that of Coma progenitors at $R\,{=}\,60$\,cMpc at redshift of 8. The result is from the FLAMINGO L1\_m8 simulation, thus being limited to $M_\ast\,{\gtrsim}\,10^8\,{\rm M_\odot}$. The medians and the 68 per cent ranges are indicated by the solid lines and the errorbars. Because it is the cumulative number of galaxies divided by the surrounding volume, the profiles become flattened at large radii. The asymptotic slopes in the inner and outer regions expected from the NFW profiles are represented by the gray dashed lines in the left panel, which are consistent with the model predictions at lower redshifts of $z\,{\lesssim}\,1$. It is notable that the number densities are remarkably similar between the protocluster types, indicating that the greater total mass of Coma progenitors is largely due to the growth in mass of member galaxies rather than their enhanced abundance. For example, the average total (stellar) mass of Coma progenitors enclosed within 5\,cMpc at $z\,{\simeq}\,3$ is about 80 (110) per cent higher than that of Fornax progenitors, whereas the number of galaxies is only ${\simeq}\,20$ per cent higher.}
\label{fig_Ngal_prof}
\end{figure*}

Aside from the total mass including dark matter, stellar mass is another fundamental quantity, which is also directly observable. Figure~\ref{fig_Ms_evol} compares the redshift evolution of total stellar mass enclosed within the varying apertures between the different types of protoclusters. As can be seen, the evolution trends observed for the total mass are also reflected in the stellar mass evolution, namely, that the protoclusters accumulate their stellar mass rapidly at $z\,{\gtrsim}\,5$, after which the mass build-up slows down to increase only gradually. In particular, unlike the evolution of the total mass of protoclusters in Fig.~\ref{fig_Mh_evol} changes its slope to be steeper around $z\,{\simeq}\,2$ compared to the earlier times of redshift between 2 and 5, the stellar mass evolution has its slope continuously decreasing with time. This may reflect quenching by feedback mechanisms effective at $z\,{\lesssim}\,2$, which makes star formation less efficient such that the rate of halo mass build-up exceeds that of forming stars. At redshifts greater than 4, the opposite trend is found that the rate at which stellar mass accumulates in both the core and outer regions of protoclusters exceeds that of dark matter. This is simply a reflection of the cosmic star formation increasing with time and peaking at $z\,{\simeq}\,2$, rather than anything specific to the environments of protoclusters. In fact, as discussed below and can be seen more clearly later (Fig.~\ref{fig_ratio_prof}), the in-situ stellar mass growth rate and star formation efficiency (SFE) are rather lower in denser protoclusters. Similarly to the previous result on the total mass, the results shown in Fig.~\ref{fig_Ms_evol} can also be utilized to identify mass and types of protoclusters as well as their $z\,{=}\,0$ fate for candidates with stellar mass and aperture known from observations, by interpolating between the different curves and panels.

\begin{figure*}
\includegraphics[width=1.\linewidth]{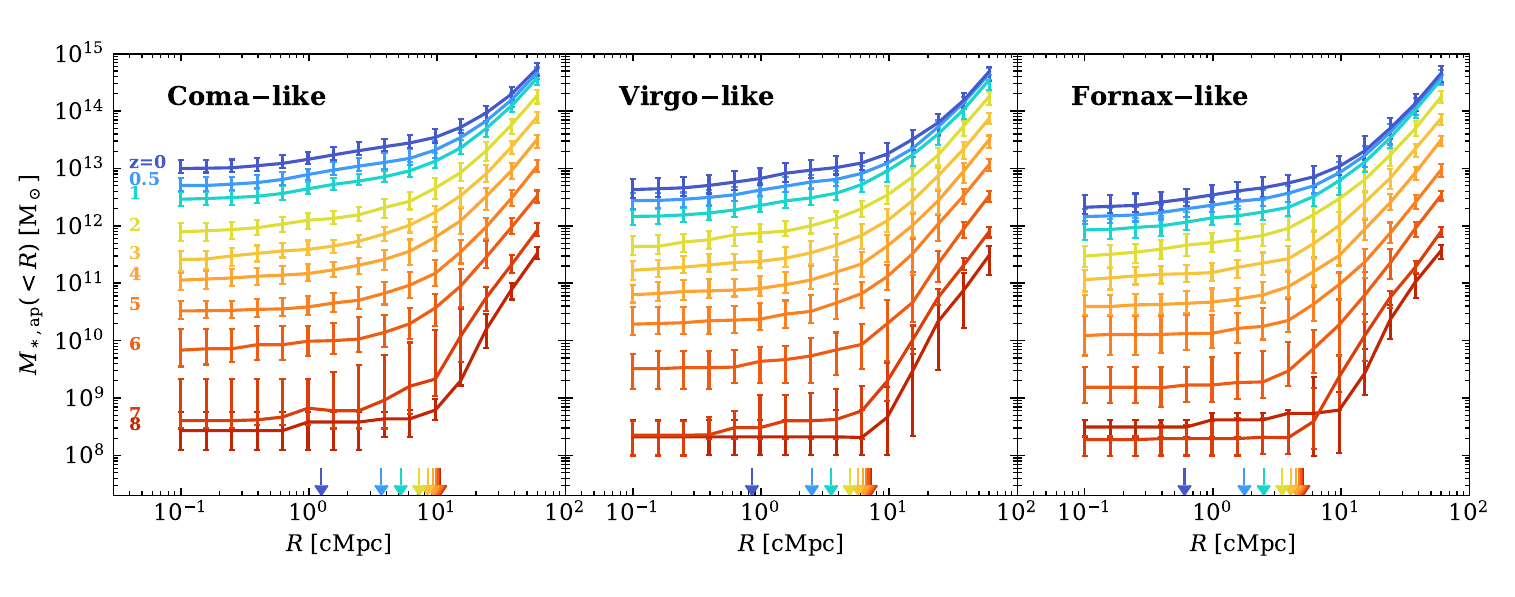}
\caption{The cumulative radial profiles of stellar mass as a function of comoving aperture and redshift for the three types of progenitors from the FLAMINGO (L1\_m8) simulation. The line styles are the same as in Fig.~\ref{fig_Mh_prof}.}
\label{fig_Ms_prof}
\end{figure*}

\begin{figure*}
\includegraphics[width=1.\linewidth]{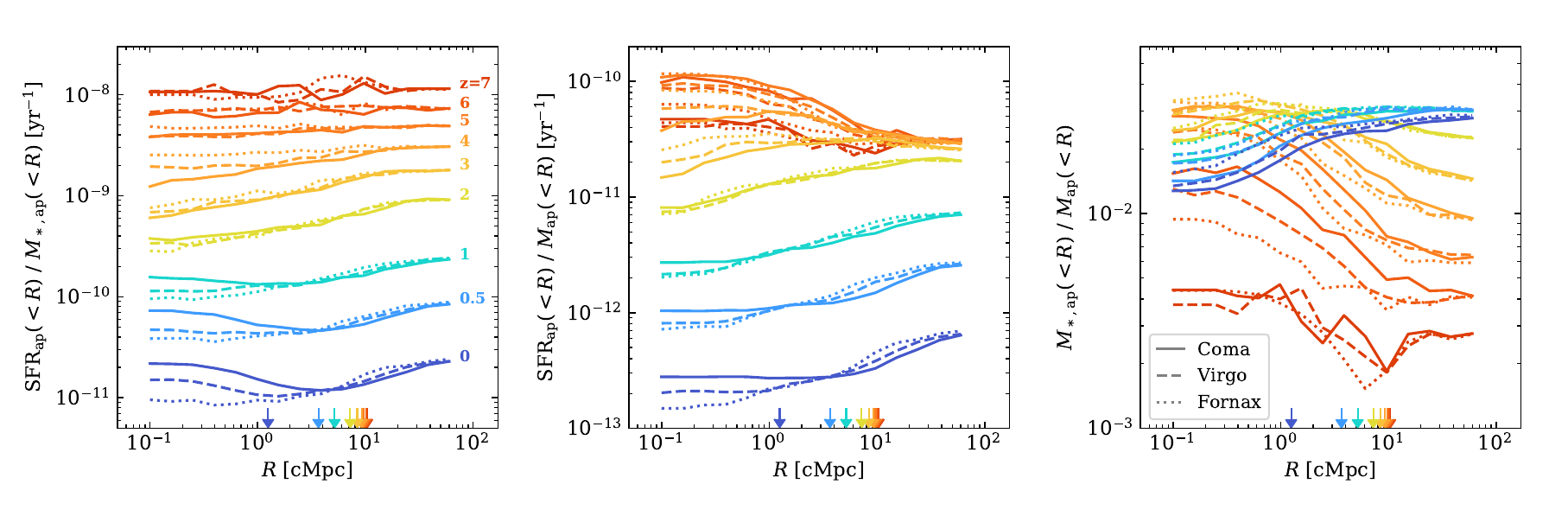}
\caption{The cumulative radial profiles of specific SFR (left panel), SFR divided by the total mass including dark matter (middle), and stellar mass-to-halo mass ratio (right) as a function of comoving aperture and redshift for Coma (solid), Virgo (dashed), and Fornax (dotted) progenitors from the FLAMINGO (L1\_m8) simulation.  The values of $R_{\rm PC}$, the sizes of typical protoclusters, for Coma progenitors, are indicated by the arrows at the bottom of each panel. }
\label{fig_ratio_prof}
\end{figure*}

\begin{figure*}
\includegraphics[width=1.\linewidth]{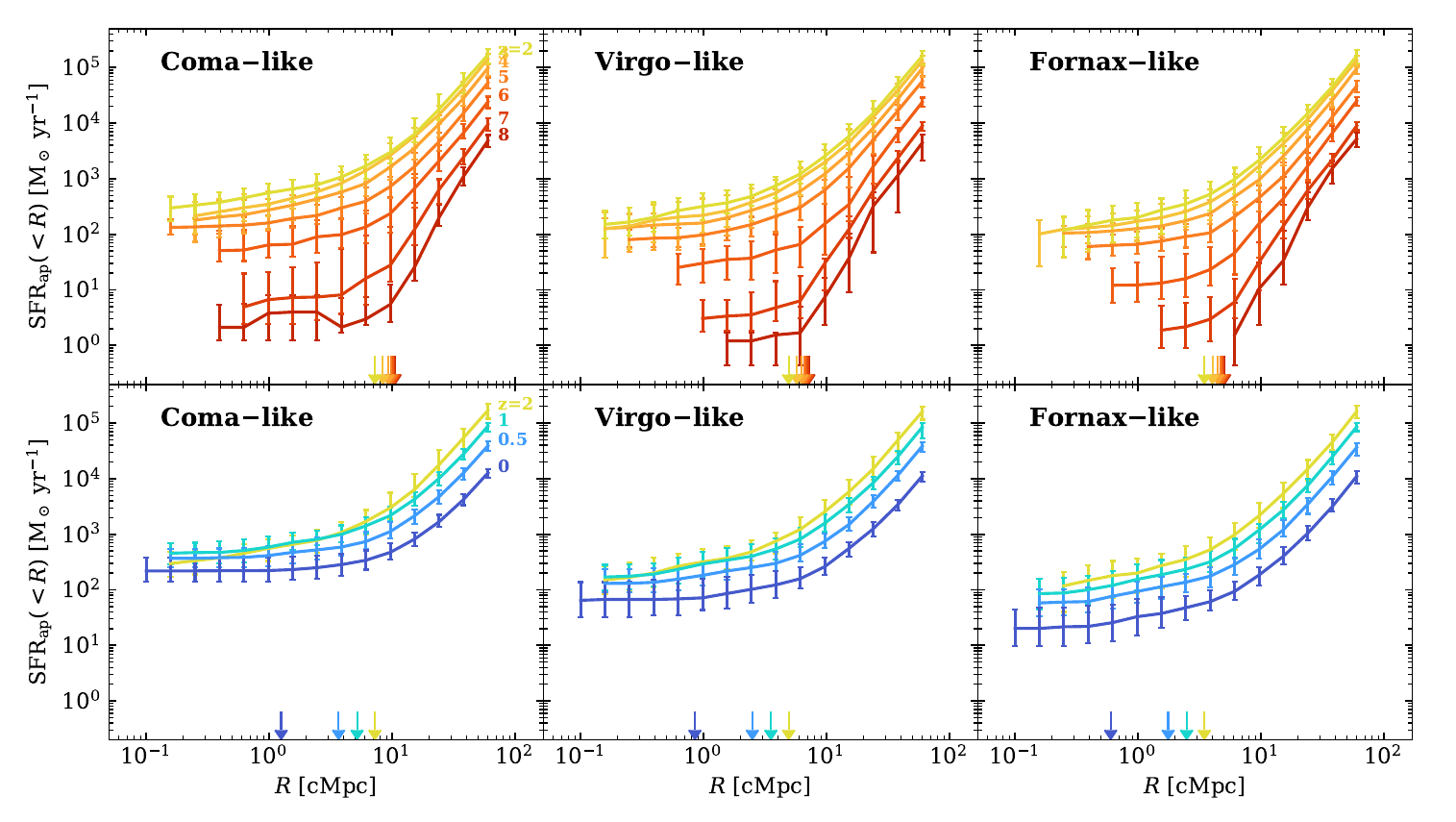}
\caption{The cumulative radial profiles of SFR as a function of comoving aperture and redshift for the three types of progenitors from the FLAMINGO (L1\_m8) simulation, separately for $z\,{\geq}\,2$ (upper panel) and $z\,{\leq}\,2$ (lower panel), for visual clarity. The line styles are the same as in Fig.~\ref{fig_Mh_prof}.}
\label{fig_SFR_prof}
\end{figure*}

\subsection{Radial profiles of physical properties}
\label{ssec_prof}

So far we have analyzed the physical properties of protoclusters and their evolution with time only within the selected apertures. Here we probe the radial profiles of protocluster properties as a function of redshift, a final piece for the complete set of information that can be combined to infer any missing quantity for protocluster candidates when only a subset of the other quantities such as redshift, aperture, aperture mass, and present-day mass (or equivalently, their $z\,{=}\,0$ fate) is available in observations. The median radial profiles of total mass and their evolution for the three types of protoclusters are shown in Fig.~\ref{fig_Mh_prof}, together with the 68 per cent ranges represented by the errorbars. The reason that the curve is flat at lower redshifts in the inner region of ${\lesssim}\,4$\,cMpc is because the total enclosed mass is dominated by the most massive halo although there are many other galaxies nearby. The mass and profiles grow by almost the same factor between the redshifts (namely, almost even spacing between the overall amplitudes of the profiles at each neighboring redshifts, which is also consistent with the more or less uniform single slope in Fig.~\ref{fig_Mh_evol}), which have a typical interval of $\Delta z\,{=}\,1$, meaning that the growth rate of matter density and structure was much greater at earlier times (as the same $\Delta z$ at higher redshifts spans a smaller amount of time). The outer profiles and slopes are determined by the cosmic density field and background galaxies that are only weakly, at maximum, associated with the central structure and do not collapse to the clusters by $z\,{=}\,0$. 

These are more directly demonstrated in Fig.~\ref{fig_Ngal_prof} where the median galaxy density profiles within the comoving spheres, normalized to that of Coma progenitors at the outermost radius of 60\,cMpc at $z\,{=}\,8$, are presented. As this probes the volume average of galaxy number density, the profiles are expected to converge to a constant at the outer radii, as is indeed shown to be the case in the plot. At the lower redshifts of $z\,{\lesssim}\,1$, the structures are shown to have developed a `core' in the central region with a shallower profile. The inner and outer profiles of structures at such stage are found to be well described by the NFW profile, which is illustrated by its asymptotic slopes shown by the gray dashed lines. Note that we only show the lines where there is more than one galaxy within the apertures. This is just for visual clarity because the centrals will only contribute to the profiles as lines with a constant slope inward through the center. Interestingly, the galaxy density profiles and their evolutions are remarkably similar between the progenitors of Coma-, Virgo-, and Fornax-like clusters, except at very low redshifts where the Coma progenitors develop slightly more established cores. This ``self-similarity'' means that the variations in the total protocluster mass within $R_{\rm PC}$, as well as those in the final cluster mass, arise from the greater volume occupied by Coma progenitors, and from the mass growth via mergers or accretion of smaller galaxies/subhaloes into bigger ones, rather than from having a more concentrated distribution of systems. 

Figure~\ref{fig_Ms_prof} probes the cumulative profiles of stellar mass within given radii as a function of redshift. The inner region is dominated by the central galaxies or protocluster cores, while the outermost region is dominated by the cosmic matter density field, approaching the same slope of ${\propto}\,r^{3}$. The overall normalization of the profiles shows a different trend of more rapid (slow) increase at high (low) redshifts compared to the total mass profiles, which reflects the evolution of star-formation rate with time as discussed above. This is directly confirmed in Fig.~\ref{fig_ratio_prof} (right panel) where the ratio of stellar mass to halo mass increases with decreasing redshift, peaks near $z\,{=}\,2$, and then decreases at the later times. 

We investigate the profiles of star-formation rate surrounding the protoclusters in Fig.~\ref{fig_SFR_prof}. Similar to Fig.~\ref{fig_Ngal_prof}, here we only show the plots where there is more than one galaxy. The results at $z\,{\geq}\,2$ and $z\,{\leq}\,2$ are separated in the upper and lower panels, respectively, for visual clarity, as the star formation turns around at the intermediate redshift. Unlike the evolution of the mass profiles, the SFR mostly `decelerates' with decreasing redshift, namely, the rate at which the SFR increases (decreases) with time becomes gradually lower (greater) with decreasing redshift, which makes the star formation per mass decrease with time in most cases, as demonstrated in the left and middle panels of Fig.~\ref{fig_ratio_prof}. An interesting thing to note is that the total SFRs within any fixed comoving volume at the high redshifts are typically within a factor of 2 between the three types of protoclusters. Given the similar galaxy number densities and much higher average total and stellar mass of individual galaxies for the Coma progenitors, we conclude that the specific SFR (sSFR) and SFE are both lower in more massive protoclusters at the high redshifts. This is seen more clearly in Fig.~\ref{fig_ratio_prof} (left and middle panels) where the star formation is shown to be less efficient in more massive protoclusters (the dotted and dashed lines being above the solid lines) at the redshift between 3 and 5. This can be due to the fact that the denser protocluster environments consist of the more massive haloes particularly in the inner region, which reach the mass regime earlier where quenching due to feedback and environmental suppression of star formation takes place. On the contrary, at the low redshifts, less massive (proto)clusters are more affected by feedback mechanisms that heat and blow the gas out of the shallower gravitational potential, resulting in more efficient quenching and less star formation, as indicated by the solid and dashed lines being above the dotted lines. At low redshifts, the innermost region in the denser environment, which is dominated by the most massive halo, appears to have the higher specific SFR than the outer region, as indicated by the `upturn' of the solid lines in the innermost region in the left panel of Fig.~\ref{fig_ratio_prof}. This counterintuitive behavior may indicate overcooling of the most massive ($M_\ast\,{\gtrsim}\,10^{12}\,{\rm M_\odot}$) galaxies in the FLAMINGO simulation, as pointed out by \citet{Schaye2023}. 

Finally, the SFR profile is quite shallow within $R_{\rm PC}$ in Fig.~\ref{fig_SFR_prof}, being dominated by the SFR in the most massive halo, at the highest redshift, which is consistent with the inside-out growth as speculated by \citet{Chiang2017}. The profile, however, becomes steeper at later times of $z\,{\lesssim}\,5$, as the SFR in the most massive haloes in the inner region is gradually suppressed by feedback in effect, while smaller haloes in the outer region reach the mass where the SFE is greatest, passing around the cosmic noon. After peaking near redshift of 2, the SFR profile slowly transitions to being flattened again through the present day, due to overall quenching and gravitational infall of member galaxies into the cluster. This `three-phase' scheme of protocluster evolution is broadly consistent with the claim of \citet{Chiang2017}, while the divides between the phases in redshift can be dependent of models and uncertainties.


\section[obs]{Comparison with observations}
\label{sec_obs}

In this section, we present comparisons of the simulation results with the properties of observed protoclusters, including the mass evolution history, star-formation history, and the number densities. We discuss some recent results from JWST in Sect.~\ref{ssec_numden}, in particular.

\subsection{Observational samples}
\label{ssec_obs_sample}

The observational samples of protoclusters we compare with the model predictions are described here, which is a compilation of data from numerous recent studies. Where the same properties of the same protoclusters have been estimated multiple times by several studies, we adopt the latest results for the comparison unless the earlier results have clear advantages over the latest. 

The first sample is a total of fourteen protocluster candidates from L21, which is a recent observational compilation of protoclusters with estimates of properties that are relatively up to date. Specifically, the sample includes four overdensities of dusty star-forming galaxies (DSFGs) from \citet{Casey2016} (the GOODS-N $z\,{=}\,1.99$ protocluster, MRC1138$-$262, SSA22, and AzTEC-3), Distant Red Core (DRC; \citealt{Oteo2018, Long2020}), and nine protocluster candidates selected from the South Pole Telescope (SPT) of SPT2349$-$56 \citep{Miller2018, Hill2020}, SPT0303$-$59, SPT0311$-$58, SPT0348$-$62, SPT0457$-$49, SPT0553$-$50, SPT2018$-$45, SPT2052$-$56, and SPT2335$-$53 (all from \citealt{WangG2021}). 

\begin{figure*}
\includegraphics[width=0.9\linewidth]{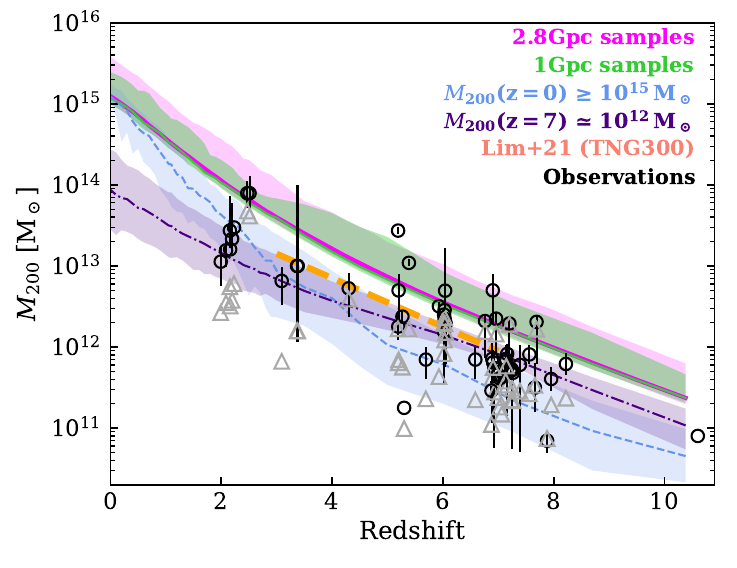}
\caption{The total mass ($M_{200}$, including dark matter) of the most massive 81 (for L1\_m8; 1679 for L2p8\_m9) protocluster cores (main haloes) identified from FLAMINGO of the large (L2p8\_m9; magenta) and fiducial (L1\_m8; green) simulation box. 81 and 1679 are the number of haloes at $z\,{=}\,0$ with the total mass greater than $10^{15}\,{\rm M_\odot}$ for each of the simulations, thus making the selection almost Coma-like in mass. For the magenta and green lines, note that this is the mass of protoclusters selected independently at each snapshot, and also the bands indicate the full range of the samples. On the other hand, the blue (purple) line indicates the median mass of progenitors (descendants) of those selected at $z\,{=}\,0$ ($z\,{\simeq}\,7$) with the total mass of $M_{200}(z{=}0)\,{\geq}\,10^{15}\,{\rm M}_\odot$ ($M_{200}(z{\simeq}7)\,{\simeq}\,10^{12}\,{\rm M}_\odot$), with the band indicating the 16-84th percentile range. A compilation of observational data is presented by the unfilled circles with the errorbars, for comparison. We adopted the observational mass estimates from the original studies, which were obtained by using stellar mass estimates and a stellar mass-to-halo mass relation from empirical models such as \citet{Behroozi2019}, and by summing up the mass within the apertures shown in Fig.~\ref{fig_aperture}. The mass corrected for the apertures, however, obtained using results similar to Fig.~\ref{fig_Mmain_Msph} but for $M_{200}$, are also presented by the gray triangles for comparison, which demonstrate the impact of the heterogenity in the observational studies and the importance of the aperture correction. Since \citet{Brinch2023} provides both the masses uncorrected and corrected for the aperture they used, we present the estimates from the original paper without further corrections. We converted to $M_{200}$ when the originally reported masses are not in $M_{200}$, but in different halo definitions such as $M_{500}$ and $M_{\rm vir}$, which is relatively negligible. While we have made no attempt to select the simulated samples to match the observations, and the observational data are heterogeneous in several aspects, the broad agreement is exhibited between the simulation predictions and observations. Finally, the orange dashed line shows the earlier results from IllustrisTNG simulation of \citet{Lim2021}, which is the average from the 25 highest-SFR galaxies at each redshift.}
\label{fig_mass_comp}
\end{figure*}

The study and samples of L21, however, are focused on SFR-based selections, identifying protocluster candidates via overdensities of DSFGs and far-IR/submm observations. As galaxies in the very early Universe are believed to be dust-poor, the discovery of protoclusters relying on DSFGs and long-wavelength observations would be limited, and the less star-forming, normal, dust-poor galaxies that emit mostly in the rest-frame optical regime could be the main tracers of protoclusters. Therefore, we also include optical-to-nearIR-selected protoclusters in the observational samples, as well as additional far-IR/submm-identified candidates recently reported in the literature. Those are ZFIRE \citep{Hung2016}, PHz G237.01$+$42.50 \citep{Polletta2021}, CC2.2 \citep{Darvish2020}, PCL1002 \citep{Casey2015}, CLJ1001 \citep{WangT2018}, all located at relatively low redshifts of between 2 and 2.5, as well as MAGAZ3NE J095924$+$022537 and MAGAZ3NE J100028$+$023349 (both from \citealt{McConachie2022}), HDF850.1 \citep{Calvi2021}, z57OD and z66OD \citep{Harikane2019}, SPT0311$-$58 as recently revisited by \citet{Arribas2023}, an overdensity at $z\,{=}\,7.66$ \citep{Laporte2022}, A2744-z7p9OD at $z\,{=}\,7.88$ \citep{Morishita2023}, and GN-z11 \citep{Tacchella2023}. Addtionally, the fifteen $z\,{\geq}\,6$ galaxy overdensities identified by \citet{Brinch2023} from the COSMOS2020 catalog \citep{Weaver2022} based on photometric redshifts using a weighted adaptive kernel are included. Finally, the seventeen protocluster candidates identified with JWST JADES, JEMS, and FRESCO at $5\,{\lesssim}\,z\,{\lesssim}\,9$ from \citet{Helton2023a, Helton2023b} described in Sect.~\ref{ssec_Msph} are compiled into the data set. 

\begin{figure*}
\includegraphics[width=0.9\linewidth]{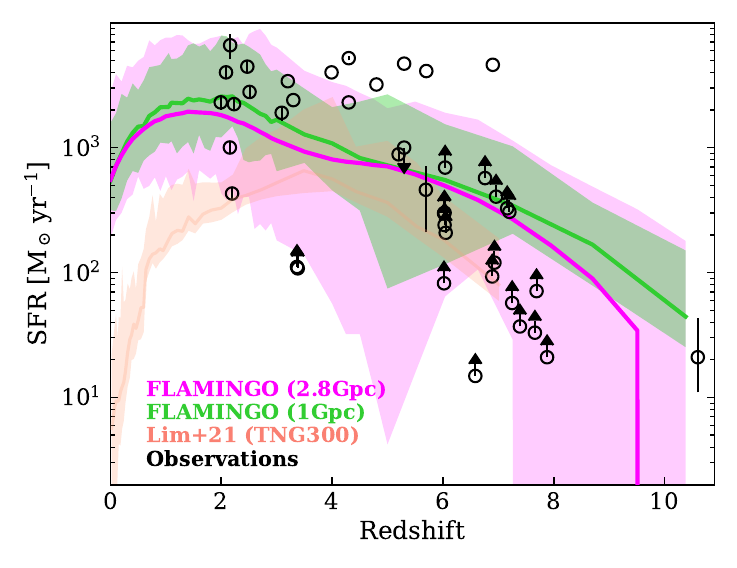}
\caption{The star-formation history (the total SFR of FoF halo) of the same mock samples as in Fig.~\ref{fig_mass_comp}. The lines indicate the medians, while the bands indicate the full range of the samples. A compilation of observational data is presented by the unfilled circles with the errorbars, for comparison. While we have made no attempt to select the simulated samples to match the observations, and the observational data are heterogeneous in several aspects, FLAMINGO predicts integrated SFR of protoclusters that are broadly consistent with the observations. For instance, some of the observational estimates do not have a well-defined aperture used to estimate the SFR in the original paper, in which case we have made no attempt to correct for their apertures. Furthermore, some of the observational tracers used to identify the candidates are expected to miss the highest-SFR members, thus a significant contribution to the total SFR, which may explain the observations falling below the predictions (also see the text). The orange curve, with the band showing the whole range of values, represents the 25 highest-SFR galaxies from IllustrisTNG of \citet{Lim2021}, which fails to reproduce the observations despite the SFR-based selection. This demonstrates an improvement in the model prediction by FLAMINGO in matching the observed SFR of the population, relative to some previous models. }
\label{fig_SFH_comp}
\end{figure*}

\subsection{Mass evolution history}
\label{ssec_MEH}

We first investigate the mass evolution history of our simulated protocluster samples, and compare it to the estimated mass of protoclusters from the observations. Most of the observations estimated the mass based on the stellar mass, using a stellar mass-to-halo mass ratio from empirical models such as \citet{Behroozi2019}. For comparisons, we select mock samples as follows. We first count the number of isolated haloes with the total mass including dark matter, $M_{200}$, greater than $10^{15}\,{\rm M_\odot}$ at $z\,{=}\,0$, thus the same as our Coma-like clusters. 81 and 1679 such objects are found from L1\_m8 and L2p8\_m9, respectively, corresponding to the number density of ${\simeq}\,0.8\times 10^7\,{\rm cMpc}^3$, roughly equal to the expected number density of Coma clusters. Then from the simulation snaphots at all redshifts, we select that number of objects as mock samples. This construction of mock samples is motivated by many of the observational studies where the most massive candidates are identified as a ``future'' Coma-like cluster. In Fig.~\ref{fig_mass_comp}, the mock samples from selected redshift snapshots of L2p8\_m9 and L1\_m8 are compared to that from the observational data. The shaded areas indicate the full range among the samples. As can be seen, the predicted mass range of simulated protoclusters reasonably covers that of the most massive high-$z$ protoclusters identified by observations. This confirms that the simulation samples are similar systems to the observed overdensities. However, when the observational data are corrected for the apertures reported in the original studies, using the calibrations shown in Figs.~\ref{fig_Mmain_Msph}, \ref{fig_Mh_evol}, and \ref{fig_Mh_prof}, the mass estimates become significantly smaller, as shown by the triangles, not predicting the evolution to the Coma-type $z\,{=}\,0$ clusters for a majority of the samples. Again, this demonstrates the impact of the heterogeneity in the observational studies, the necessity of the aperture correction, and how critical our results in Figs.~\ref{fig_Mh_evol}--\ref{fig_SFR_prof} are for facilitating the corrections. The previous results based on IllustrisTNG simulations from L21, selecting the 25 systems with the highest SFR, are also shown for comparison, which exhibits another factor of about 1.5 decrease in mass relative to L1\_m8, mainly due to their smaller box size than the FLAMINGO suites. We confirmed that when the selection is made based on the FoF SFR, the median predictions are barely affected while the scatter becomes greater to be about 0.3\,dex, which is as expected because the selection is now not directly based on the halo mass. 

We also explored the mass evolution history of Coma-like clusters, namely those with the total mass greater than $10^{15}\,{\rm M}_\odot$ at $z\,{=}\,0$, as well as of those with the total mass similar to $10^{12}\,{\rm M}_\odot$ at $z\,{\simeq}\,7$ which roughly matches the mass of the candidates from the observations near the redshift. These two cases are indicated by the blue and purple lines, respectively, in Fig.~\ref{fig_mass_comp}, with the bands representing the 16-84th percentiles. It is seen that the mass history of the Coma-like clusters has a wide range of scatter for individual objects that spans an order of magnitude at $z\,{\gtrsim}\,3$. Similarly, not all massive objects of $M_{200}\,{\simeq}\,10^{12}\,{\rm M}_\odot$ at $z\,{\simeq}\,7$ end up being a massive cluster by redshift of zero. Both results suggest a great uncertainty in predicting the fate of high-$z$ objects solely based on their mass. Also, the median trends show that the most massive $z\,{=}\,0$ clusters rather have the smaller mass at the high redshifts than those that end up at the lower $z\,{=}\,0$ mass. In part, this reflects the late assembly of
massive objects in the structure growth predicted by $\Lambda$CDM cosmology. This complicates speculations on the true fate of protocluster candidates from high-$z$ observations. 

The fact that some of the observational data points (particularly after the aperture correction) fall below the path predicted to evolve to a Coma-like cluster by $z\,{=}\,0$ is as expected, given the number density of the Coma cluster in the local Universe, less than one per $10^7\,{\rm cMpc}^3$, compared to the relatively small volumes probed so far by the high-$z$ observations. Instead, according to the model predictions, the less massive protocluster candidates from the observations would become Virgo/Fornax-like clusters or massive groups at $z\,{=}\,0$. This will be also demonstrated in Sect.~\ref{ssec_numden}.

\begin{figure*}
\includegraphics[width=1.\linewidth]{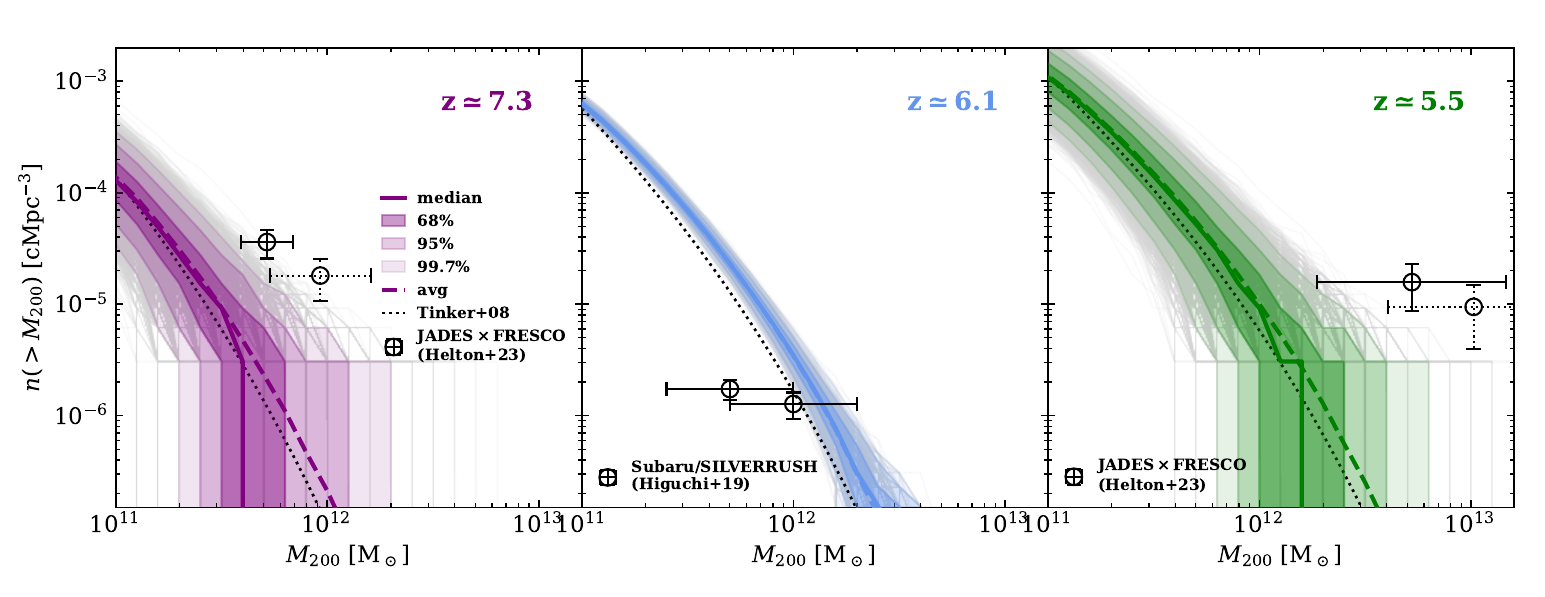}
\caption{The cumulative number density of protocluster cores (the most massive haloes) as a function of $M_{200}$ at three selected redshifts, calculated from each of subboxes of L2p8\_m9 with the volume corresponding to that of the observations, to account for the cosmic variance. The size of the subboxes is about 70\,cMpc for the comparisons at $z\,{=}\,7.3$ and 5.5, while 230\,cMpc at $z\,{=}\,6.1$. The results from each subbox are shown by individual lines in gray, with the shaded bands showing 68, 95, and 99.7 per cent ranges, while the median and total average are represented by the solid and dashed lines. For reference, the halo mass function of \citet{Tinker2008} at each redshift is also presented by the dotted lines. The recent estimates with JWST observations based on \citet{Helton2023b}, shown by the circles with the errorbars and bands in the left and right panels, are shown to be significantly higher than the theoretical predictions. Helton et al. obtained the halo mass estimates based on the UniverseMachine of \citet{Behroozi2019}. The number densities in case of some of the samples in the close proximity considered as a single object to merge later, are shown by the circles with the dotted errorbars. The results based on the Subaru/SILVERRUSH survey from \citet{Higuchi2019}, on the other hand, shown in the middle panel, present a much better agreement with the models. Because Higuchi et al. estimated the halo mass based on the galaxy overdensity, only accounting for the most massive halo within the whole protocluster region, this hints at the importance of correcting for the aperture size within which the physical quantities of protoclusters are calculated. }
\label{fig_numden}
\end{figure*}

\subsection{Star-formation history}
\label{ssec_SFH}

Because protocluster are expected to contribute significantly to the cosmic star-formation history as demonstrated earlier in Sect.~\ref{ssec_basic} and Fig.~\ref{fig_CSFRD}, it is important to make sure that the SFRs of individual protoclusters from the simulation reproduce those from the observations. This is shown in Fig.~\ref{fig_SFH_comp}, where the median and full range of integrated SFR of the simulated protoclusters from L1\_m8 and L2p8\_m9 are compared to the subset of the observational samples for which the SFR estimates are available from the literature. The simulation samples selected here are the same as in Sect.~\ref{ssec_MEH}. We confirm that there is no significant change in the SFR of the cores from the simulations when using the SFR within different apertures to select samples. For reference, the earlier results from IllustrisTNG by L21 are also presented with the median and 68 per cent range. As shown and discussed in L21, IllustrisTNG is found to underpredict the total SFR by up to a factor of 10 compared to the observations, even though the results shown here are for the TNG samples selected based on the SFR, namely from 25 protoclusters with the highest SFR at each snapshots. The star-formation history predicted for the FLAMINGO samples, particularly those from the largest box, however, are seen to match the observational data much better. The improvement is found to be most remarkable near redshifts of 2 to 3 where the observational samples and estimates are most robust. 

Unlike the mass evolution history, we haven't done uniform aperture corrections for the observational SFR except for some of them. This is partly because not all of the observations report a well-defined aperture in the original study. However, a more fundamental reason is that the tracers used by the observations are often times expected to be biased towards identifying only a subset of true member galaxies according to their SFR, making it extremely difficult to assess the uncertainties in each of the estimates. The observations with very high SFRs of 3000 to $5000\,{\rm M_\odot\,yr^{-1}}$ between redshifts of 4 and 6, in particular, which are mostly the SPT-selected candidates from \citet{WangG2021}, need to be further investigated in detail (see also \citealt{Brinch2023}). Wang et al. identified the bright unresolved SPT sources (fields) as potential candidates, and followed them up with the Atacama Pathfinder Experiment (APEX) telescope’s Large APEX BOlometer CAmera (LABOCA; \citealt{Kreysa2003, Siringo2009}) to resolve individual sources within each of those candidate fields. Their SFR estimates, however, were derived via SED fitting using a single photometry of LABOCA at 870\,$\micron$. The uncertainties also involved in their assumed dust temperature and conversion from the far-infrared luminosity to SFR, in addition to that in the SED fitting, make their SFR estimates less robust. Furthermore, not all of their sources have confirmation of membership with spectroscopic redshifts, potentially leading to an over-estimation of the integrated SFR of their protocluster candidates, although we already adopted the lower limit provided in their paper which only accounts for their high signal-to-noise sources. Finally, the aperture within which they computed the integrated SFR, which ranges from 5 to 12\,cMpc, is typically much larger than that of our results or other observations. On the other hand, some observational data, particularly those with the uparrows should be considered as lower limits, as those samples are mostly optically-selected galaxies only, potentially missing associated SMGs or DSFGs. That may explain the apparent discrepancy of SFR from the simulation at the high redshifts of 6 to 8 relative to the data.

\subsection{Number densities and cosmic variance}
\label{ssec_numden}

So far, only dozens of protoclusters have been identified in the observations, and most of them were identified individually e.g. by using rare types of galaxies including radio galaxies and SMGs as signposts to search around them with follow-up observations to reveal normal galaxies in the neighborhood. It was not until recently that studies sought and identified protocluster candidates systematically within a fixed survey volume, enabling estimates of the number densities of protoclusters at high redshifts. One of such recent observations is by \citet{Helton2023a,Helton2023b} who combined the JWST JADES, JEMS, and FRESCO surveys to identify a total of seventeen protoclusters with a large number of spectroscopically confirmed member galaxies, as described earlier in Sect.~\ref{ssec_Msph}. Given the overlapping area of about $81\,{\rm arcmin}^2$ between the surveys that were used for their study, and assuming the cosmology we adopt throughout our analysis, the corresponding comoving volume from which their samples are selected using H$\alpha$ and [OIII]$\lambda 5008$ detections is approximately $3.2\times 10^5\,{\rm cMpc}^3$ and $3.3\times 10^5\,{\rm cMpc}^3$, spanning $4.9\,{<}\,z\,{<}\,6.6$ and $6.7\,{<}z\,{<}\,8.9$, respectively. The halo masses were inferred from the mass of objects with the properties matching the observations, using the empirical model of \citet{Behroozi2019}. Another example is the study by \citet{Higuchi2019} who discovered 14 and 26 protocluster candidates from the LAE samples of the Systematic
Identification of LAEs for Visible Exploration and Reionization Research Using Subaru HSC (SILVERRUSH; \citealt{Ouchi2018}) around $z\,{\simeq}\,5.7$ and 6.6 over the sky area of 13.8 and $16.2\,{\rm deg}^2$, respectively. Given the effective survey volumes computed by \citet{Higuchi2019} of $1.1\times 10^7\,{\rm cMpc}^3$ and $1.5\times 10^7\,{\rm cMpc}^3$ for each of the two redshift ranges, the number density of protocluster candidates is estimated to be $1.2\times 10^{-6}\,{\rm cMpc}^{-3}$ and $1.7\times 10^{-6}\,{\rm cMpc}^{-3}$, respectively. They estimated the halo mass of the candidates by obtaining the average mass of the haloes with the similar overdensities from the theoretical model of \citet{Inoue2018}. 

To compare the number density of the observed protoclusters with the simulation prediction, we take into account the sample variance expected for the survey volumes corresponding to the data. Specifically, we divide the whole simulation box into sub-volumes whose size each matches the survey volumes. We use the results from L2p8\_m9 as it is the simulations containing the largest number of sub-volumes of 64,000 (1,728) subboxes matching the survey volume of \citet{Helton2023b} \citep{Harikane2019}, which is about 70 (230)\,cMpc, thus most suitable to explore the sample variance with high statistical significance. The comparisons are made based on the simulation snapshots with redshifts close to the redshift range of each observational data. 

Figure~\ref{fig_numden} shows the cumulative number density of protoclusters identified from the subboxes, together with the median (solid curves), 68 (dark-shaded), 95 (shaded), and 99.7 (lightly shaded) per cent ranges to demonstrate the cosmic variance. The average from the whole simulation box of L2p8\_m9 is provided as the dashed curve, which is also compared to the theoretical halo mass function of \citet{Tinker2008} for reference. The mass estimates for the JWST samples adopted from \citet{Helton2023b} were obtained by summing up the mass of haloes associated with galaxies within their apertures of choice used to define protoclusters. They obtained each halo mass based on the UniverseMachine of \citet{Behroozi2019} that applied a semi-empirical model constrained by numerous observations including SMFs, SFRs, quenched fractions, luminosity functions, and auto- and cross-correlations of galaxies. That is probably the reason why it greatly exceeds the model prediction by about a factor of ten, or at about 3-sigma tension, at both $z\,{\simeq}\,7.3$ and 5.5, because our results for the model prediction only account for the mass of most massive haloes, as theoretical studies normally do including \citet{Chiang2017}. Based on Fig.~\ref{fig_Mmain_Msph}, the mass correction to account for the apertures used by Helton et al. is indeed expected to be a factor of 2 to 4, which is roughly the right amount to shift the data points to the left and resolve the apparent tension. While the UniverseMachine used a different definition of halo mass, the impact on the results is negligible relative to the other factors. Higuchi et al., on the other hand, provide the mass already accounting only for the most massive haloes of their protocluster candidates. The fact that it shows greater agreement with our results, thus reassures that if the choice of aperture for mass estimates is properly accounted for, the mass function of protoclusters from observations agrees with the theoretical predictions. Some early data from JWST claimed discovery of galaxies and their abundances at high redshifts that may not be explained well by the current paradigm of cosmology and galaxy formation (e.g. \citealt{Labbe2023, Boylan-Kolchin2023}). It is worth mentioning that, unlike the stellar mass function or some other observables for which the model predictions may vary due to the uncertainties in the baryonic physical processes, the halo mass function is predicted more robustly with great accuracy and precision if the assumed cosmology is correct. 

About a half of the Helton et al. samples are along the similar LOSs and in the close proximity kinematically to each other, indicating that some of them might merge later to form a single (proto)cluster instead of multiples. In such case, their actual number densities would be lower, roughly by a factor of 2. However, the individual and average mass of the samples also increase through the mergers, shifting the number densities to the right in Fig.~\ref{fig_numden} as well. As a result, such consideration preserves the discrepancy at a similar level, as indicated by the dotted points.

\section[discussion]{Discussion}
\label{sec_discussion}

\subsection{Impact of numerical resolution}
\label{ssec_numres}

While the FLAMINGO simulation suite has the great advantage of large box sizes to contain extreme protoclusters or environments reported from observations, such a large box size was achieved at the expense of numerical resolution of $m_{\rm gas}=1.34\times10^8\,{\rm M_\odot}$. Given the lower stellar mass limit of protocluster galaxies of approximately $2\times10^7\,{\rm M_\odot}$ from recent observations at the high redshifts considered in this study \citep[e.g.,][]{Helton2023a, Helton2023b}, the lower resolution of the simulation may raise a concern (although the observational samples are suspected to be not complete down to the mass limit). Depending on the fraction of contribution from those unresolved low-mass haloes, the physical properties of ovedensities we investigated might be biased potentially significantly in the worst scenario. As an example, the amount of aperture correction as investigated in Figs.~\ref{fig_Mmain_Msph}--\ref{fig_Ms_evol} might be under-estimated compared to the true values, considering that the lower-mass haloes are less biased tracers of the matter density field, and thus their relative contribution to the integrated quantities are expected to be greater in the outer, less dense regions. Similarly, the limited resolution may impact the radial profiles explored in Sect.~\ref{ssec_prof}, given the general dependence of galaxy clustering on their properties including mass. We plan to investigate this in detail in future work, using zoom simulations that go down to about two orders of magnitude better mass resolution than L1\_m8 (private communication with Doug Rennehan). On the other hand, the results on the number density of main progenitor (the most massive halo, or the protocluster `core') and their redshift evolution will not be affected by the numerical resolution for the objects in the mass range of interest. 

As an approximate approach, we fitted the cumulative contribution to the quantities such as SFR, by galaxies as a function of stellar mass, using a Schechter function, and then extrapolated the function to the lower mass down to $10^7\,{\rm M_\odot}$. This simple consideration estimates the further contribution from the unresolved galaxies would be up to a factor of 2 and much smaller in most cases. The missing contribution increases with increasing redshifts, however. We also reach a similar amount of the missing contribution when comparing with the results from L1\_m9, namely the FLAMINGO suites with an order of magnitude lower mass resolution.

\begin{figure*}
\includegraphics[width=0.9\linewidth]{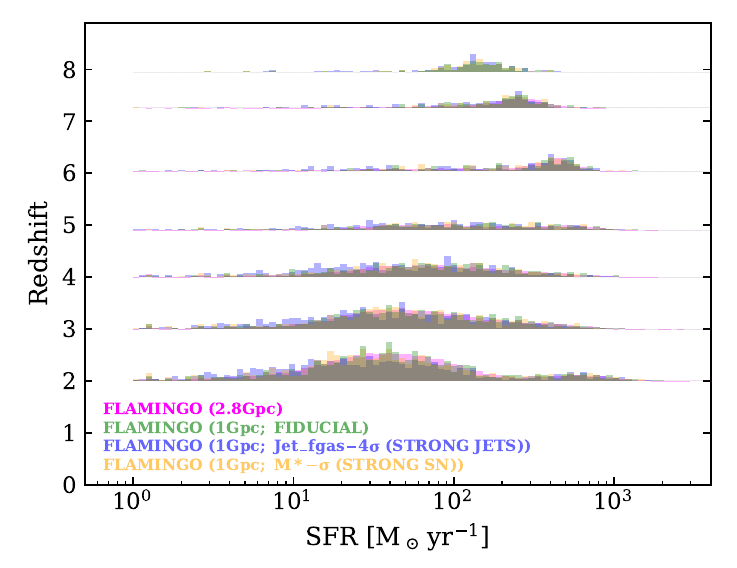}
\caption{The SFR of individual member galaxies in the same mock protocluster samples as in Fig.~\ref{fig_mass_comp}, and the impact of subgrid physics models on the predictions (histograms). Only the results from two selected model variations showing the strongest changes relative to the fiducial model, which are ``${\rm Jet_{-}fgas}$\,--\,$4\sigma$'' (strong jets; blue) and ``M*\,--\,$\sigma$'' (strong SN feedback; yellow) are shown. The predictions from varying models do not show significant differences with each other, indicating that the uncertainties involved in the assumed subgrid models can be neglected given the current uncertainties in models and observations. Although no attempt was made for the selection of mock samples to match the observations, FLAMINGO predicts galaxies with about a thousand solar mass per year at $3{\lesssim}\,z\,{\lesssim}\,5$ as in the observations, unlike some previous theoretical models \citep{Lim2021}. By investigating a subbox of the simulation matching the size of e.g. TNG, we confirm that such improvement for matching the observations is largely due to the increased volume of the simulation. }
\label{fig_SFR_ind}
\end{figure*}

\subsection{Impact of subgrid models and cosmology}
\label{ssec_subgrid}

Finally, we employ the variations of L1\_m9 to explore the impact of assumed subgrid models. As described in Sect.~\ref{ssec_FLAMINGO}, the FLAMINGO suites have a total of eight simulation runs where changes in the parameters responsible for the strengths of stellar and AGN feedback are implemented to test the uncertainties in their fiducial physics model. However, the parameters and models are varied only to a degree such that the model prediction still remains compatible with the observed stellar mass function and cluster gas fraction at $z\,{\simeq}\,0$ within reasonably assumed systematics of 2 to 8\,$\sigma$. 

In Fig.~\ref{fig_SFR_ind}, we present the model predictions from selected L1\_m9 variations. The simulation samples are the same as in Sect.~\ref{ssec_MEH} and \ref{ssec_SFH}, namely the one hundred protocluster cores with the highest SFR in each snapshot. The model predictions with the various subgrid models are more or less consistent with each other, indicating that our results are insensitive to details of physics models. While we present the results from only two selected models of ``${\rm Jet_{-}fgas}$\,--\,$4\sigma$'' (strong jets) and ``M*\,--\,$\sigma$'' (strong SN feedback), which are confirmed to produce the most significant changes with respect to the fiducial model, the other models exhibit even smaller changes in their predictions. Also, while we only show the results for the individual SFR as an example, similar conclusions of only minor changes in the predictions are found for the other results too. This perhaps surprising similarity between the models may be due to the process how the model parameters are tuned in the FLAMINGO project. Similar to when changing the resolutions, the model parameters are re-adjusted to still match the other observations as they change a subset of parameters in the subgrid models. Namely, the ${\rm Jet_{-}fgas}$\,--\,$4\sigma$ model has their parameters tuned to match the $z\,{=}\,0$ SMF while calibrated to the lower cluster gas fraction by $4\sigma$. Another thing to note is that the FLAMINGO models predict galaxies with high SFR of about a thousand solar mass per year at $z\,{\simeq}\,3$--5, matching that of the SMGs and DSFGs from observations, which some earlier results from simulations with smaller box size have been lacking (see L21). 

The cosmology variations of FLAMINGO suites, including the Planck with higher neutrino mass and the lower $S_8$ as motivated by observations of the low-redshift Universe (often referred to as the $S_8$ tension), are only found to reinforce the tension by predicting the smaller number densities of overdense systems. The Planck cosmology, on the other hand, predicts slightly more protocluster cores, although the change is negligible relative to the amount of the discrepancy to mitigate the tension at all. The cosmology variations do not affect either the rest of our analysis and conclusions on the radial profiles and redshift evolution of properties. The protocluster abundance drops typically by slightly less than a factor of ${\simeq}\,$3 and 2, almost constantly over the mass and redshift probed in our analysis, in the low $S_8$ and high neutrino mass variations, respectively. The tension against \citet{Helton2023b} at $z\,{\simeq}\,7.3$ becomes as large as 3.6\,$\sigma$ when based on the low $S_8$ simulation.

\section[summary]{Summary and conclusions}
\label{sec_summary}

In this study, we examined the properties of mock protocluster samples predicted from the FLAMINGO simulation with an observationally motivated identification. An unprecedented advantage of FLAMINGO simulation is its large box size of up to ${\simeq}\,22\,{\rm cGpc}^3$, or 2.8\,cGpc on a side, as protoclusters are the rarest objects or environments whose number density is believed to be about $10^{-7}\,{\rm cMpc}^{-3}$, with some extreme subsets having even lower number densities down to an order of $10^{-8}\,{\rm cMpc}^{-3}$ \citep[e.g.,][]{Miller2015, Casey2016, Rennehan2020}. Our analysis and findings can be summarized as follows. \newline

1. The large FLAMINGO simulation suite allowed us to conduct a first reliable statistical analysis of the population. In observations, the identification of protoclusters and the prediction of their fate at $z\,{\simeq}\,0$ are usually carried out by matching the estimated total mass associated with identified member galaxies to the theoretical prediction. To mimic such observational selection most closely, we first used the simulation merger tree to track the true progenitors of massive $z\,{=}\,0$ clusters at high redshifts. Then we select all haloes of mass equal to the average mass of the most massive true progenitors at a given redshift, as potential candidates. Among those candidates, we only consider those without more massive haloes within $R_{90}(z)$, the average comoving 3D radius that encloses 90 per cent of $z\,{=}\,0$ members at each redshift, as protocluster samples. We divided our samples into three types of progenitors of Coma-, Virgo-, and Fornax-like clusters, according to the mass of the main progenitor halo, or `core' as normally referred to as in observations. The mass criteria for Fornax-, Virgo-, and Coma-like clusters are $(1\,{-}\,3)\times 10^{14}\,{\rm M_\odot}$, $(3\,{-}\,10)\times 10^{14}\,{\rm M_\odot}$, and ${>}\,10^{15}\,{\rm M_\odot}$ in FoF mass at $z\,{=}\,0$, respectively. \newline

2. We found that most of the most massive haloes from the simulation with the average mass matching that from the merger tree of protoclusters are isolated, namely, having no bigger haloes in their neighborhood, and thus expected to evolve into clusters at later times. Also, the simulated protoclusters, with the three types combined, exhibit a contribution of more than 20 (50) per cent to the CSFRD at redshifts greater than 2 (4), consistent with the findings from previous studies \citep[e.g.][]{Chiang2017}, confirming their importance for scrutinizing the evolution of galaxies and the early Universe (Fig.~\ref{fig_CSFRD}). At later times of $z\,{\lesssim}\,3$, Virgo and Fornax progenitors, rather than the Coma progenitors, provide a major contribution to the CSFRD because of their higher number density. On the contrary, at the high redshifts of $z\,{\gtrsim}\,5$ including the EoR, the CSFRD is found to be dominated by the progenitors of Coma-like clusters. However, more work is required to confirm a quantitative assessment of their fractional contribution to the reionization, as the FLAMINGO suites do not possess a mass resolution high enough to resolve low-mass haloes and examine their contribution during the EoR thoroughly. \newline

3. Among the heterogeneities in studies of protoclusters are the various apertures adopted in the literature as seen in Fig.~\ref{fig_aperture}, a complication that has largely been neglected in previous studies. Understanding the impact of using different apertures for the same protoclusters is one of the main goals of our study. We find that the impact of the choice of aperture is indeed significant, easily resulting in a factor of 2 to 3 bias in the estimates of total mass of progenitors at $z\,{\gtrsim}\,5$, and increasing with time to more than an order of magnitude at $z\,{\lesssim}\,4$ for the apertures typically assumed for observational estimations (see Fig.~\ref{fig_Mmain_Msph}). This, therefore, can potentially lead to an overestimation of their future mass evolution and a misidentification of protoclusters if no correction for the aperture is made. \newline

4. We also inspected the redshift evolution of total mass and stellar mass of protoclusters within select comoving volumes in Figs.~\ref{fig_Mh_evol} and \ref{fig_Ms_evol}. We found that the progenitors of Coma-like clusters begin with haloes of ${\simeq}\,10^{11}\,{\rm M_\odot}$, on average, at $z\,{\simeq}\,8$, and increase their mass rapidly by accreting mass until $z\,{\simeq}\,3$ after which they become condensed gradually to collapse and form clusters. Similar evolutionary trends are found for the Virgo and Fornax progenitors except at lower mass scales. The stellar mass growth of protoclusters, on the other hand, is found to slow down at the later time of $z\,{\leq}\,2$ due to quenching by feedback mechanisms, while increasing more quickly at the higher redshifts compared with the evolution of the total mass. \newline 

5. Our investigation of radial profiles revealed that the total masses of protoclusters is dominated by the haloes associated with the central galaxies. At later times of $z\,{\lesssim}\,1$ when the (proto)clusters are found to have developed a ``core'', the profiles of galaxy number densities match NFW profiles (Fig.~\ref{fig_Ngal_prof}). The number density profiles demonstrate a notable similarity between the different types of protoclusters, i.e. being ``self-similar'', indicating that their mass contrast is mainly due to that in the mass growth of individual member galaxies, via mergers and accretion, instead of having more galaxies at given moments. The profiles of total stellar mass and integrated SFR are dominated by the most massive haloes in the inner region in the early Universe of $z\,{\gtrsim}\,5$, and then become steeper at later times of redshift between 5 and 2 where the central haloes undergo gradual quenching while the smaller nearby haloes peak in the SFE approaching the cosmic noon. After $z\,{\simeq}\,2$, due to overall quenching and infall of member galaxies into the final cluster, the dominance by the inner haloes is recovered. \newline

6. In Sect.~\ref{sec_obs}, we made comparisons of our results with observations. A compilation of observational data we used consists of a set of protoclusters confirmed and well-studied in the literature, as well as some newly discovered candidates including those from the recent JWST observations by \citet{Helton2023a, Helton2023b}. Both dusty starburst galaxies selected in the far-IR/submm such as DSFGs and SMGs, and `normal' galaxies selected mainly in the near-IR or rest-frame optical bands are represented in the compiled samples. In these comparisons, our mock samples were constructed by choosing the same number of the most massive objects from each snapshot as that of objects at $z\,{=}\,0$ with the total mass greater than $10^{15}\,{\rm M_\odot}$, to make the selection thus Coma-like in mass. Unlike some results from earlier studies using other simulations, the predictions of total SFR from the FLAMINGO project are found to be in reasonable agreement with the observational estimates (Fig.~\ref{fig_SFH_comp}), particularly considering the potentially large systematics and uncertainties, heterogeneous selection techniques relying on a variety of tracers, and lack of our careful attempt to mimic the observational selections. \newline

7. The number density of protoclusters estimated from recent systematic searches within fixed survey volumes is not always consistent with the theoretical expectations. Even when taking into account the cosmic variance relevant for the survey volumes, we find that some studies show a discrepancy of greater than 3\,$\sigma$ with the number density being higher by an order of magnitude than our simulation predictions, while other studies are consistent with the prediction (Fig.~\ref{fig_numden}). Given our results above about the impact of apertures for estimating the total mass, we conclude that the disagreement can be attributed mainly to an overestimation of mass for the observed protoclusters, as the tension is greatly diminished when the correction for the apertures are accounted for. \newline

8. One concern about the results from the FLAMINGO simulation might be its mass resolution not matching the lower mass limit of observed galaxies that are used to trace the overdensities, which go down to approximately $2\times10^7\,{\rm M_\odot}$ in stellar mass. The simulation may be potentially missing the contribution from the lower-mass objects for some of the predictions. We find that the SFR and stellar mass increase by up to a factor of 2 when the mass resolution is improved by a factor of 8. However, the subgrid parameters of the FLAMINGO simulations have been tuned to match the $z\,{=}\,0$ SMF up to $M_\ast\,{=}\,10^{11.5}\,{\rm M_\odot}$ and low-$z$ cluster gas fraction at each resolution. \newline

9. Finally, we probed the model predictions from various simulation runs where different parameters for stellar and AGN feedback were assumed, in order to evaluate uncertainties associated with the assumptions on subgrid physics. The model predictions are consistent with each other, thus showing no hints of significant concerns related to the assumptions on uncertain astrophysical processes (Fig.~\ref{fig_SFR_ind}). \newline

In this study, we investigated the evolution and profiles of protoclusters mainly based on the mass-based selection, particularly focusing on the impact of apertures. Another major uncertainty, however, is a variety of techniques and tracers used for the identification of protoclusters. This will be the focus of our future work.

\section*{ACKNOWLEDGEMENTS}

We thank the anonymous referee for the constructive comments. SL and RM acknowledge support by the Science and Technology Facilities Council (STFC) and by the UKRI Frontier Research grant RISEandFALL. RM also acknowledges funding from a research professorship from the Royal Society. This work used the DiRAC@Durham facility managed by the Institute for Computational Cosmology on behalf of the STFC DiRAC HPC Facility (\url{www.dirac.ac.uk}). The equipment was funded by BEIS capital funding via STFC capital grants ST/K00042X/1, ST/P002293/1, ST/R002371/1 and ST/S002502/1, Durham University and STFC operations grant ST/R000832/1. DiRAC is part of the National e-Infrastructure. This work is partly funded by research programme Athena 184.034.002 from the Dutch Research Council (NWO).

\section*{DATA AVAILABILITY}

The data underlying this article will be shared on reasonable request to the corresponding author.

\bibliographystyle{mnras}
\bibliography{protocluster2.bib}



\label{lastpage}

\end{document}